\documentclass[twoside,twocolumn,9pt]{article}
\usepackage{extsizes}
\usepackage[super,sort&compress,comma]{natbib} 
\usepackage[left=1.5cm, right=1.5cm, top=1.785cm, bottom=2.0cm]{geometry}
\usepackage{balance,enumitem}
\usepackage{sectsty}
\usepackage{graphicx} 
\usepackage{lastpage}
\usepackage{float}
\usepackage{fancyhdr}
\usepackage{fnpos}
\usepackage[english]{babel}
\addto{\captionsenglish}{%
  
}
\usepackage{array}
\usepackage[symbol]{footmisc}

\usepackage{charter}
\usepackage[T1]{fontenc}
\usepackage[usenames,dvipsnames]{xcolor}
\usepackage{setspace}
\usepackage[compact]{titlesec}
\usepackage{hyperref}
\usepackage{bm}
\definecolor{cream}{RGB}{222,217,201}
\usepackage{changes}

\usepackage{amsmath}
\usepackage{amsfonts}
\usepackage{amssymb}
\usepackage{todonotes}
\usepackage{float}
\usepackage{widetext}

\def\bea{\begin{eqnarray}}
\def\eea{\end{eqnarray}}
\def\la{\langle}
\def\ra{\rangle}

\def\l{\left}
\def\r{\right}

\def\bl{\begin{align}}
\def\el{\end{align}}

\def\bea{\begin{eqnarray}}
\def\eea{\end{eqnarray}}
\def\ba{\begin{array}}
\def\ea{\end{array}}
\def\n{\nonumber}
\def\c{\mathscr}

\def\be{\begin{equation}}
\def\ee{\end{equation}}
\def\bse{\begin{subequations}}
\def\ese{\end{subequations}}

\def\dr{D_R}
\def\de{D_{\text{eff}}}
\def\o{\omega}
\def\t{\theta}
\def\sig{\sigma}

\def\g{\gamma}
\def\s{\sigma}

\def\lm{\lambda}
\def\n{\nonumber}

\newcommand{\eref}[1]{Eq.~(\ref{#1})}%
\newcommand{\fref}[1]{Fig.~\ref{#1}} %
\newcommand{\sref}[1]{Sec.~\ref{#1}}%
\newcommand{\aref}[1]{Appendix~\ref{#1}}%

\newcommand{\ie}{{i.e., }}

\begin{document}

\pagestyle{fancy}
\thispagestyle{plain}

\twocolumn[
  \begin{@twocolumnfalse}

{\huge\textbf{Chirality Reversing Active Brownian Motion in Two Dimensions}}

\noindent\large{Santanu Das,$\dag$ \textit{$^{a,b}$} and Urna Basu, \textit{$^{c}$} } \\
 
\noindent $^{a}$ School of Physical Sciences, National Institute of Science Education and Research, Jatni 752050, India \\
\noindent $^{b}$Homi Bhabha National Institute, Training School Complex, Anushakti Nagar 400094, India \\
\noindent $^{c}$S N Bose National Centre for Basic Sciences, Kolkata 700106, India \\

{\bf Abstract}\\
\noindent We study the dynamics of a chirality reversing active Brownian particle, which models the chirality reversing active motion common in many microorganisms and microswimmers. We show that, for such a motion, the presence of the two time-scales set by the chirality reversing rate $\gamma$ and rotational diffusion constant $\dr$ gives rise to four dynamical regimes, namely, (I) $t \ll \text{min}(\g^{-1}, \dr^{-1})$, (II) $\g^{-1} \ll t \ll \dr^{-1}$, (III) $\dr^{-1} \ll t \ll \g^{-1}$ and (IV) $t \gg \text{max}(\g^{-1}, \dr^{-1})$, each showing different behaviour. The short-time regime (I) is characterized by a strongly anisotropic and non-Gaussian position distribution, which crosses over to a diffusive Gaussian behaviour in the long-time regime (IV) via an intermediate regime (II) or (III), depending on the relative strength of $\gamma$ and $\dr$. In regime (II), the chirality reversing active Brownian motion reduces to that of an ordinary active Brownian particle, with an effective rotation diffusion coefficient which depends on the angular velocity. Finally, we find that, the regime (III) is characterized by an effective chiral active Brownian motion.

\end{@twocolumnfalse} \vspace{0.6cm}

]

\section{Introduction}

Active particles are self-propelled units that are able to consume energy from the surrounding medium and convert it into a persistent motion \cite{bechinger2016active, marchetti2013hydrodynamics,
ramaswamy2017active, gompper20202020, jiang2010active, romanczuk2012active, solon2015active, howse2007self}. Examples of such active motion are observed in various physical contexts, ranging from biological systems, soft-matter systems to granular systems. Broadly, active motion can be categorized into two classes, linear and chiral.  Linear active particles move via self-propulsion but without any self-rotation, the trajectories showing no specific `handedness'. Chiral active motion, on the other hand, refers to the scenario when the active particles self-rotate with an angular velocity, in addition to self-propulsion. The different directions of the self-rotation and self-propulsion leading to typically circular or helical trajectories~\cite{sevilla2016diffusion, van2008dynamics, kummel2013circular, kruger2016curling, liebchen2022chiral}. Physically, such active chiral motions may originate due to a variety of reasons, including asymmetry in shape~\cite{kummel2013circular, wykes2016dynamic} or propulsion mechanism~\cite{nosrati2015two, diluzio2005escherichia}, and hydrodynamic coupling with interfaces~\cite{lauga2006swimming}.

It has recently been observed that an active particle can reverse its chirality in time~\cite{narinder2018memory, crenshaw1996new,ordemann2003pattern, takagi2013dispersion, van2009clockwise,lancia2019reorientation,  ebata2015swimming,tarama2011dynamics,hokmabad2019topological}. Reversing the chirality implies flipping a clockwise circular or helical motion to a counter-clockwise motion. For example, in response to light, unicellular green algae \emph{Chlamydomonas reinhardtii} \cite{crenshaw1996new} and zooplankton \emph{Daphnia} \cite{ordemann2003pattern} can reverse the direction of their chiral motions. Another example is a spherical light-activated colloidal microswimmer in a viscoelastic fluid that can reverse its sense of rotation above a critical propulsion speed \cite{narinder2018memory}.  Chemically propelled rods, even with a slight curvature in their shape, yield two stable states of curved orbits between which they can stochastically switch due to the thermal fluctuations \cite{takagi2013dispersion}. Circular swimmers in petri-dish and ring-like confinements flip their orientational sense of motion in time \cite{van2009clockwise}. Reversal of chirality is also observed in soft active objects like swimming droplets driven by a surface wave \cite{ebata2015swimming}, light-responsive spiral droplets \cite{lancia2019reorientation}, deformable self-propelled particles under external field \cite{tarama2011dynamics}, and self-propelling nematic shells \cite{hokmabad2019topological}.

In spite of a widespread appearance, theoretical understanding of chirality reversing motion is quite limited, even at the single particle level. Most of the previous studies in this context have been focused on the effective diffusive behaviour at late times, for both stochastic and deterministic reversal of chirality~\cite{haeggqwist2008hopping, weber2011active, weber2012active,olsen2021diffusion,shenoy2007kinematic}. In fact, even the short-time  behaviour of the mean-squared displacement of the particle, which carries prominent signatures of activity~\cite{bechinger2016active}, is not known yet. Moreover, no theoretical results are available for the higher order position fluctuations of the chirality reversing active particles.

%


In this paper, we study the chirality reversing active motion in two dimensions using a simple model of a chiral active Brownian particle which stochastically reverses its direction of self-rotation with a constant rate. The presence of the chirality reversal rate $\gamma$ introduces an additional time-scale $\gamma^{-1}$, along with the typical activity time-scale $\dr^{-1}$, the inverse of the rotational diffusion constant $\dr.$ 
We show that the presence of these two time-scales gives rise to four distinct dynamical regimes: (I) $t \ll \min(\gamma^{-1}, \dr^{-1})$, (II) $ \gamma^{-1} \ll t \ll \dr^{-1}$, (III) $ \dr^{-1} \ll t \ll \gamma^{-1}$, and (IV) $t \gg \max(\gamma^{-1}, \dr^{-1})$, each characterized by distinctly different dynamical behaviour. In the short-time regime (I), the position distribution of the chirality reversing active Brownian particle (CRABP) is strongly anisotropic and non-diffusive, which we compute analytically. This crosses over to an isotropic Gaussian distribution in the late time regime (IV) via an intermediate regime (II) or (III), depending on whether $\dr$ is smaller or larger than $\gamma.$ We show that, in the intermediate regime (II) the dynamical behaviour of CRABP is similar to an ordinary active Brownian particle (ABP) with an effective rotational diffusion constant. On the other hand, in the intermediate regime (III), the CRABP dynamics resembles that of a purely chiral active Brownian particle (CABP).

The paper is organized as follows. In the following section, we define the model and present a brief summary of our main results.  In \sref{mean_var_pos} we compute the first two moments of the position exactly and show their different characteristics in different dynamical regimes. The position distributions in the four distinct regimes are computed in \sref{pdf_pos}. We conclude with some general remarks in \sref{conclusion}.

\section{Model and Results} 
\label{model}

We consider a chiral active Brownian particle moving with a constant speed $v_0$ in two dimensions, which is described by its position $(x,y)$ and orientation $\theta(t)$ at time $t$. The orientation evolves via rotational diffusion, and a self-rotation with an angular velocity, the sign of which determines the chirality of the particle. The chirality reversal is introduced by allowing the reversal of the sign of the angular velocity with a constant rate. The time evolution of position and orientation of such a chirality reversing active Brownian particle is then described by the Langevin equations,
\bse
\bea
\label{vxt}
\dot{x}(t) &=& v_0 \cos\theta(t),\\ 
\label{vyt}
\dot{y}(t) &=& v_0 \sin\theta(t),\\ 
\dot{\theta}(t) &=& \omega \sigma(t) +  \sqrt{2 D_R} \; \eta(t),
\label{theta_t}
\eea
\label{eq:model}
\ese
where $\omega$ denotes the angular speed,  $\dr$ denotes the rotational diffusion constant and $\eta(t)$  is a white noise with $\la \eta(t) \ra = 0$ and  $\la \eta(t) \eta(t')\ra = \delta(t - t')$. The dichotomous noise $\sigma(t)$ alternates between $\pm 1$ with a constant rate $\gamma$ triggering the reversal of the chirality.  It has an exponentially decaying autocorrelation,
$\la \sig(t) \sig(t')\ra = e^{-2 \g |t - t'|}.$

\begin{figure}[t]
\centering \includegraphics[width= 0.9\hsize]{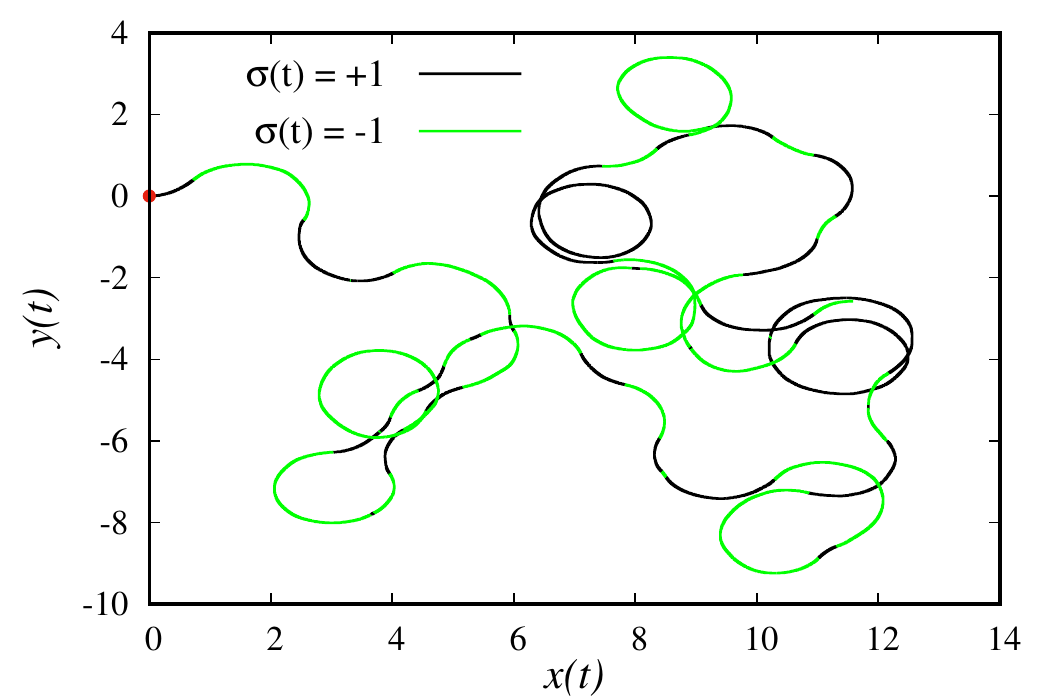}~
\caption{\label{traj} A typical trajectory of a CRABP, starting from the origin [marked by the red dot],  up to time $t = 100$ for a chirality reversing rate $\g = 0.5$ with  fixed $v_0 = 1 = \o$ and rotational diffusion constant $D_R = 0.01$. Black and green lines indicate the instantaneous signs of chirality $\sigma(t)=+1$ and $\sigma(t)=-1$, respectively.} 
\end{figure}

Note that the time evolution of the orientation $\theta(t)$, governed by \eref{theta_t}, is identical to the dynamics of the position of a run and tumble particle (RTP) in one dimension \cite{malakar2018steady}. The angular velocity $\o \sig(t)$ in our model plays the role of the velocity of a RTP that can flip its sign at a constant rate $\g$. In the absence of flipping of $\sig(t)$ \ie for $\g = 0$, the $\theta$  dynamics reduces to that of a CABP~\cite{van2008dynamics}. On the other hand, for $\o = 0$, \eref{eq:model} describes an ordinary ABP \cite{basu2018active}. 

The above model for chirality reversing active motion was introduced recently, in the context of studying the swimming trajectories of several biological and synthetic active matter systems where it was shown that at late times the particle has an effective diffusive behaviour~\cite{olsen2021diffusion}. However, no analytical results beyond the late-time effective diffusion constant are known for this model. 

In the following, we comprehensively characterize the effect of the reversal of angular velocity on the position fluctuation of the chiral active particle. Without any loss of generality, we consider the initial condition $x=0=y$, along with $\theta=0$ i.e., the particle starts from the origin, oriented along the $x$-axis. 
We also consider that initially, the chirality of the particle can be positive or negative with equal probability, i.e., at $t=0$, $\sigma= \pm 1$ with probability $1/2$. Fig.~\ref{traj} shows a typical trajectory of the CRABP where the two different colours indicate the instantaneous sign of the chirality. Before going to the details of the computation, we present a brief summary of our main results first.

\begin{figure*}[t]
\centering\includegraphics[width= 0.99\hsize]{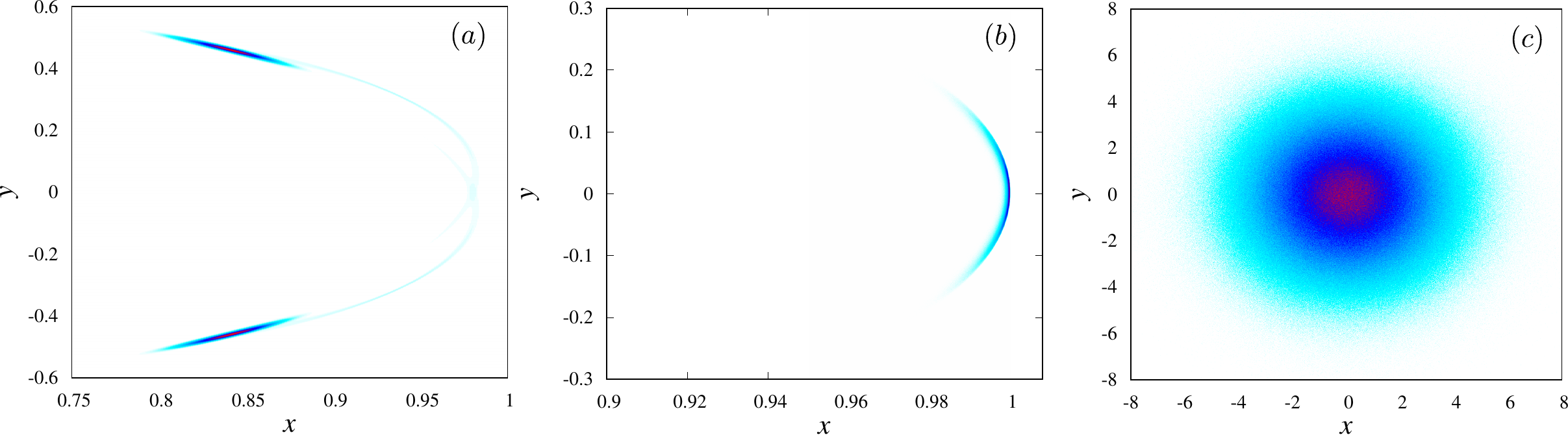}~
\caption{\label{crabp_2d} Position distribution of a CRABP in regime (I), (II) and (III) are plotted in (a), (b) and (c) respectively. We consider $\dr = 10^{-3}$, $\g = 0.1$ in (a), $\dr = 10^{-3}$, $\g = 10^2$ in (b) and $\dr = 20$, $\g = 10^{-3}$ in (c) at time $t = 1$, $1$ and $80$ respectively, with $v_0 = 1 = \o$. Darker regimes in the diagram correspond to higher probability.}
\end{figure*}

The CRABP has two intrinsic timescales --- $\dr^{-1}$ and $\g^{-1}$, set by the orientational diffusivity and the reversal rate of angular velocity, respectively. Evidently, the first time-scale, $\dr^{-1}$, characterizes the persistence of the orientational angle whereas $\g^{-1}$ is the typical time between consecutive reversal events. We show that the presence of these two time-scales gives rise to four distinct and dynamically different regimes. We find both the short- and late-time regimes where $t \ll \min(\dr^{-1}, \g^{-1})$ and $t \gg \max(\dr^{-1}, \g^{-1})$ respectively, with an intermediate regime where $\g^{-1} \ll t \ll \dr^{-1}$ or $\dr^{-1} \ll t \ll \g^{-1}$  depending on whether $\g$ is larger or smaller than $\dr$.


To understand the nature of the position fluctuations in the different regimes, we first compute the mean and the variance of the $x$ and $y$ position components exactly [see Eqs. \eqref{mu_xt}, \eqref{mu_yt}, \eqref{mnt_xa}, \eqref{mnt_ya}]. We find that when $\gamma < \omega$, i.e., when the chirality reversal rate is smaller than the angular speed, both moments show oscillatory behaviours (with different periods) in the intermediate time regime--- the amplitude of the oscillations decays exponentially, and an effective diffusive behaviour comes out at late times. We also go beyond the first two moments, and explore the marginal position distributions  by considering the effective Langevin equations in the different dynamical regimes:

\begin{itemize}[leftmargin=*]

\item \textbf{Short-time regime (I)} [$t \ll \min(\g^{-1}, \dr^{-1})$]: In this short-time regime, the CRABP has a strongly anisotropic and non-diffusive motion. The anisotropy is clearly visible in Fig.~\ref{crabp_2d} (a) which shows a plot of the $P(x,y,t)$ in this regime. The anisotropy and non-diffusive nature of the motion are more apparent from the behaviour of the variance of the position --- the variance of the $x$ component of the position $\sigma_x^2(t) \sim t^4$ whereas  $\sigma_y^2(t) \sim t^3$ [see Eqs. \eqref{varx_st} and \eqref{vary_st}]. 

We use a perturbative approach to compute the marginal distributions $P(x,t)$ and $P(y,t)$ in this regime [see Eqs. \eqref{eq:Px_shortt} and \eqref{eq:Py_shortt}] which turn out to have very different and rather unusual qualitative features. The $x$-marginal distribution $P(x,t)$ is strongly skewed [see Fig.~\ref{crabp_2d} (a)] with a Gaussian peak near $x = v_0 t$ and an additional `shoulder' closer to $x = v_0 t$, originated due the presence of the chirality reversal. On the other hand, the $y$-marginal distribution $P(y,t)$ is  symmetric and double-peaked, with a plateau-like region near the origin [see Fig.~\ref{pdf_small_t}].


\item \textbf{Intermediate regime (II)} [$\g^{-1} \ll t \ll \dr^{-1}$]: We show that in this intermediate regime the CRABP dynamics is similar to the dynamics of an ordinary ABP with an effective rotational diffusion constant,
\bea
D_A = D_R + \frac{\o^2}{2 \g}.
\label{DA}
\eea
Depending on values of $\o t$ in this regime we observe two different behaviours of the effective ABP.
When $\o t \sim O(1)$ we find that $t \ll D_A^{-1}$ where CRABP exhibits short-time behaviour of ABP. In this regime, the $x$-distribution has a non-trivial scaling form  [\eref{Fz}] as a function of the scaled variable $(v_0 t-x)/v_0 D_A t^2$, whereas $y$ distribution is Gaussian with a width $\sim t^3$. Evidently, the anisotropic and non-diffusive nature of the motion also survives in this regime [Fig.~\ref{crabp_2d}(b)]. In contrast to this anisotropy, an isotropic Gaussian distribution of position with the variance $\sig_x^2(t) = \sig_y^2(t) \simeq v_0^2 t/D_A$ obtained in the limit $\o t \gg 1$ where $t \gg D_A^{-1}$.

\item \textbf{Intermediate regime (III)} [$\dr^{-1} \ll t \ll \g^{-1}$]: This intermediate regime is characterized by an isotropic and diffusive motion, similar to a passive Brownian particle with an effective diffusion constant independent of $\gamma$,
\bea
D_\text{CABP} = \frac {v_0^2}2 \frac {\dr}{\dr^2 + \o^2}.
\label{D_CABP}
\eea
Correspondingly, the typical fluctuations along both $x$ and $y$ directions are described by Gaussian distributions with width $D_\text{CABP} t$ [see Figs.~\ref{crabp_2d}(c) and \ref{pdf_3}]. In fact, we show that, this diffusive behaviour is same as that of an ordinary CABP, with angular speed $\omega$ and rotational diffusivity $\dr$, in its long-time regime $ t \gg \dr^{-1}$.

\item \textbf{Late-time regime (IV)} [$t \gg \max(\g^{-1}, \dr^{-1})$]: In this long-time regime also the typical position fluctuations of CRABP is isotropic and diffusive, both $P(x,t)$ and $P(y,t)$ being Gaussian in nature [see \eref{var_reg4} and Fig.~\ref{pdf_4}]. However, the effective diffusion constant in this regime depends on both $\dr$ and $\gamma$,
\bea
\de = \frac{v_0^2}{2} \l( \frac{\dr + 2\g}{\o^2 + \dr( 2 \g + \dr)} \r).
\label{d_eff}
\eea 
We observe this $\de$ a non-monotonic function of $\dr$ for a given $\g$, whereas it increases monotonically with $\g$ when $\dr \to 0$ and becomes $\g$ independent when $\dr \to \infty$ [see inset of \fref{var_crabp}(b)].
\end{itemize}

The emergence of the different dynamical regimes due to the presence of multiple active time-scales has previously been observed in the context of direction reversing active Brownian particles \cite{santra2021active}. In that case, the reversal of the propulsion direction gives rise to the additional time-scale, which makes the behaviour of the emergent dynamical regimes very different than what we observe here with chirality reversals.


In the next sections we discuss the behaviour of the different dynamical regimes in details.

\section{Moments of position}
\label{mean_var_pos}

To get an idea about the nature of the position fluctuations, we first compute the mean and variance of the $x$ and $y$ components. As already mentioned, we assume that the particle starts from the origin $x=y=0$ at $t=0$. Then, from \eref{eq:model}, it is straightforward to write,
\bea
\la x(t) \ra  =  \int_0^t ds \, \la \xi_x(s) \ra, ~~\text{and}~~ \la y(t) \ra  =  \int_0^t ds \, \la \xi_y(s) \ra, \label{eq:mean}
\eea
where, for notational simplicity, we have denoted the active noises,
\bea
\xi_x(t)  \equiv v_0 \cos \t (t), ~~~~\text{and}~~~~ 
\xi_y(t)  \equiv v_0 \sin \t (t).
\label{eff_noise}
\eea
From \eref{eq:model}, the second moments of position can similarly be expressed 
as
\begin{subequations}
\bea
\la x^2(t) \ra  =  \int_0^t ds \int_0^t ds' \, \la \xi_x(s) \xi_x(s') \ra, \\
\la y^2(t) \ra  =  \int_0^t ds \int_0^t ds' \, \la \xi_y(s) \xi_y(s') \ra.
\eea
\label{eq:x2y2}
\end{subequations}
Clearly, to explicitly compute the position moments, we need the mean and auto-correlations of the active noise which can be computed by using the Fokker-Planck equation corresponding to Eq.~\eqref{theta_t}. Using the explicit forms of the noise correlations, we can calculate the moments of the position. The details of these computations are given in \aref{effec_noise}.

The mean displacement of the CRABP along $x$-axis is given by [see \aref{appendix_mnts} for the details], 
\bea 
\mu_x(t) &=& \frac{v_0 (2 \g + \dr)}{\l(\dr^2 + 2 \g \dr + \o ^2\r)} \l[  \l( 1 - e^{-t (\g + \dr)} \cosh \l( \lm_1 t \r) \r) \r. \cr
&& +\l.   \l( \frac{\o^2}{(2 \g + \dr)} -\g \r) e^{-t (\g + \dr)} \frac{\sinh \l( \lm_1 t \r)}{\lm_1} \r], \label{mu_xt}\\ 
\mu_y(t) &=& 0 \label{mu_yt} 
\eea
where $\lm_1 = \sqrt{\g^2 - \o^2}$. Here we have used the initial condition $\theta(0)=0$ and $\sigma(0)=\pm 1$ with equal probability $1/2$. 

From Eq.~\eqref{mu_xt} we see that the parameter $\lm_1$ controls the qualitative behaviour of $\mu_x(t)$. For $\g < \o$, \ie when $\lm_1 < 0$, $\mu_x(t)$ exhibits an oscillatory behaviour. The amplitude of this oscillation decays exponentially and $\mu_x(t)$ saturates to a constant value $\frac{v_0 (2 \g + \dr)}{\l(\dr^2 + 2 \g \dr + \o ^2\r)}$ for $t \gg (\g + \dr)^{-1}$. On the other hand, for $\g \ge \o$ (\ie when  $\lm_1 \ge 0$), there is no oscillation in $\mu_x(t)$ -- it increases monotonically and eventually reaches the constant value for 
$t \gg (\g + \dr - \sqrt{\g^2 - \o^2})^{-1}$. Note that, using $\g = 0$ and $\o = 0$ in the above equation we get back the results of $\mu_x(t)$ for a CABP \cite{van2008dynamics} and an ABP \cite{basu2018active} respectively.

\begin{figure*}[t]
\centering{
\includegraphics[width=0.85\hsize]{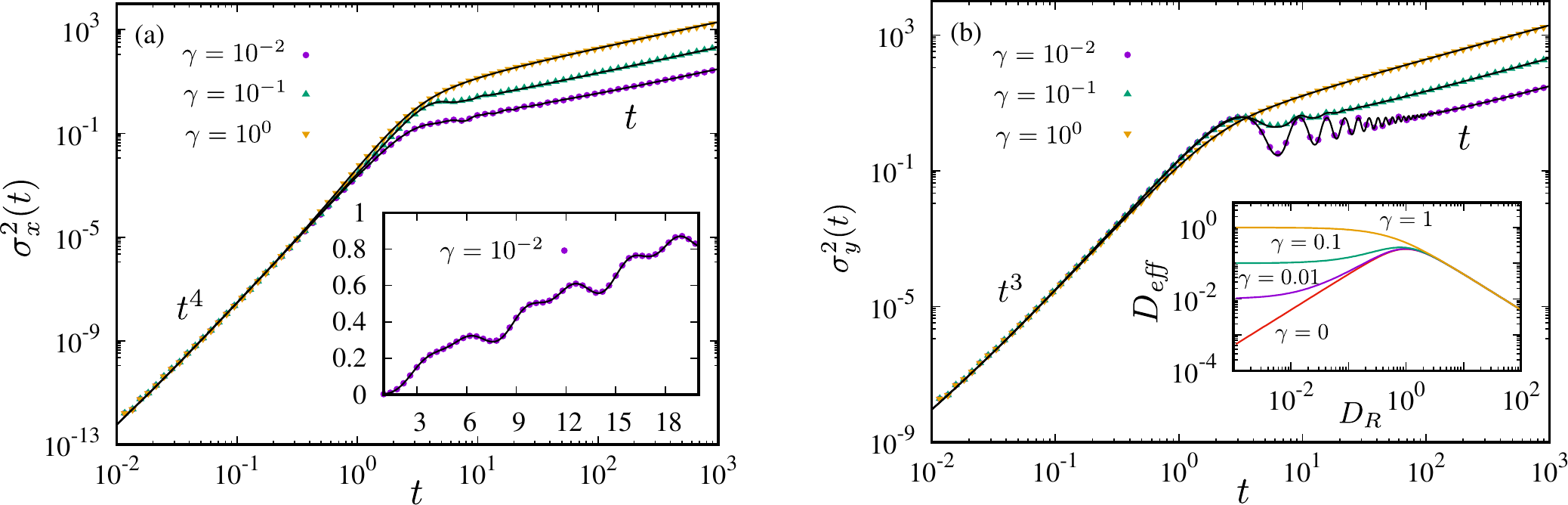}~
\caption{\label{var_crabp} Variance of $x$ and $y$ components of position of a CRABP are plotted in (a) and (b) respectively as functions of time $t$ for different values of chirality reversion rate $\g$ with the fixed $v_0 = 1$, $\omega = 1$, $D_R = 0.01$. Symbols denote the data obtained from numerical simulation and black solid lines denote the analytical predictions [See Eqs. \eqref{varxy}, \eqref{mu_xt}, \eqref{mu_yt}, \eqref{mnt_xa} and \eqref{mnt_ya}].  Inset (a): we highlight the oscillatory behaviour of $\sig_x^2(t)$ in the limit $\lm_1 < 0$. Inset (b): non-monotonic behaviour of $\de$ with respect to $\dr$ is plotted for different $\g$ by taking equal values of all the other variables (\ie $v_0$, $\o$ and $\dr$) as the main plot.}}
\end{figure*}

Next we discuss the behaviour of the variance of $x$ and $y$ components which are, by definition,
\bea
\sig_x^2(t) = \la x^2(t) \ra - \mu^2_x(t), ~~\text{and}~~
\sig_y^2(t) = \la y^2(t) \ra - \mu^2_y(t). 
\label{varxy}
\eea
The second moments $\la x^2(t) \ra$ and $\la y^2(t) \ra$ can be computed using Eq.~\eqref{eq:x2y2} along with Eqs. \eqref{cor_xxa} and \eqref{cor_yya} ; see \aref{effec_noise} for the details. The corresponding explicit expressions are rather long and are given by Eqs.~\eqref{mnt_xa} and \eqref{mnt_ya} in the Appendix. Using them in \eref{varxy} alongside $\mu_x(t)$ and $\mu_y(t)$ from Eqs. \eqref{mu_xt} and \eqref{mu_yt} respectively, we extract analytical results for the variance that are plotted in Fig.~\ref{var_crabp} along with the data from numerical simulations. Here we observe the variance exhibits an exponentially-decaying oscillatory behaviour for $\lm_2 = \sqrt{\g^2 - 4\o^2} < 0$. This oscillatory behaviour completely dies out when $\lm_2 \ge 0$. We illustrate this behaviour in Fig.~\ref{var_crabp}. 

It is also interesting to look at the second moment of the radial distance $r$, $\la r^2 \ra = \la x^2 \ra + \la y^2 \ra$, which can be expressed in a simple analytical form,
\begin{align}
\la r^2(t) \ra =& 4 \de \,t +  \frac{2 v_0^2\big[(\dr + 2 \g)^2 - \o^2 \big]}{(\dr(\dr + 2 \g) + \o^2)^2} \l[e^{-(\dr+\g)t} \cosh(\lm_1 t) - 1 \r]~~ \cr
+ &\frac{2 v_0^2}{\lm_1} \l[ \frac{\g (\dr + 2 \g)^2 - (2\dr + 3 \g) \o^2}{(\dr(\dr + 2 \g) + \o^2)^2}\r] e^{-(\dr+\g)t} \sinh(\lm_1 t). 
\label{mnt2_r}
\end{align}
Once again, we get back the results of CABP \cite{van2008dynamics, ebbens2010self} and ABP \cite{bechinger2016active} by taking $\g = 0$ and $\o =0$ respectively in the above equation. 

Now we briefly mention the qualitative behaviour of the position variance in the different dynamical regimes. \\

\noindent{\bf Short-time regime (I):} In this regime $t \ll \min(\g^{-1}, \dr^{-1})$, and taking $e^{-\g t} \simeq 1-\g t$ and  $e^{-\dr t} \simeq 1-\dr t$, as $\g t \ll 1$ and $\dr  t\ll 1$ respectively, we extract 
\noindent
\bea
\n
\label{varx_st}
\sig_x^2(t) &\simeq& 
\frac{1}{3} \dr^2 v_0^2 t^4  - \frac{1}{15} \dr v_0^2 t^5 \l(7 \dr^2 - 4 \o^2 \r)  \\ &+&\frac{1}{90} \dr v_0^2 t^6 \l(35 \dr^3 - 51 \dr \o^2 - 13 \g \o^2 \r), \\ \n
\sig_y^2(t) &\simeq& \frac{2}{3} \dr v_0^2 t^3 + \frac{1}{12} v_0^2 t^4 \left(3 \o^2 - 10 \dr^2 \r) \\  &+& \frac{1}{30}  v_0^2 t^5 \left(21 \dr^3 - 19 \dr \o^2 - 4 \g \o^2 \r)
\label{vary_st}
\eea
by using Eqs.\eqref{varxy}, \eqref{mu_xt}, \eqref{mu_yt}, \eqref{mnt_xa}, and \eqref{mnt_ya}. Clearly the variance in the $x$ direction $\sig_x^2(t) \sim t^4$ is much smaller than that in the $y$ direction $\sig_y^2(t) \sim t^3$, which we find consistent with the result of the same for an ABP \citep{basu2018active}. The effect of chirality and the corresponding reversal rate $\g$ appear in the subleading order terms $\sim t^5$ and $\sim t^6$  for the $x$ and $\sim t^4$ and $\sim t^5$ for the $y$ components of the variance respectively. All these differences along $x$ and $y$ directions clearly reflect an anisotropy on the distribution of position in this regime. This anisotropy is originated from the breaking of the symmetry of dynamics along $x$ and $y$ directions due to the initial consideration of $\t_0 = 0$. \\

\noindent{\bf Intermediate regime (II):} In this regime, where $\gamma t \gg 1 $, while $\dr t \ll 1$, the behaviour of the variance can show two subtly different scenarios, depending on whether $\omega$ is larger than $\gamma$ or not. For $\o < \gamma$, using $e^{-\g t} \simeq 0$, we get, from Eqs.~\eqref{mnt_xa} and \eqref{mnt_ya} [see \aref{int_reg1} for the details], 
\begin{align}
\label{var_eabp_reg2x}
\sigma_x^2(t) &\simeq \frac{v_0^2 t}{D_A} + \frac{v_0^2}{12 D_A^2}\;\l( e^{-4 D_A t} - 12 e^{-2 D_A t} + 32 \;e^{- D_A t} - 21 \r),\\
\sigma_y^2(t) &\simeq \frac{v_0^2 t}{D_A} - \frac{v_0^2}{12 D_A^2}\;\l( e^{-4 D_A t} - 16 \;e^{- D_A t} + 15\r),
\label{var_eabp_reg2y}
\end{align}
where $D_A$ is given by \eref{DA}. The above equations represent the variance on an ordinary ABP with rotational diffusion constant $D_A$, with distinctive short-time and long-time behaviours\cite{basu2018active}. In particular, at times $t \ll D_A^{-1}$, it shows an anisotropic $\sim t^3$ growth, which crosses over to a diffusive $\sim t$ behaviour for $t \gg  D_A^{-1}$. On the other hand, for $\omega > \gamma$, we get from Eqs.~\eqref{mnt_xa} and \eqref{mnt_ya},
\bea
\sigma_x^2(t) = \sigma_y^2(t) \simeq  \frac{v_0^2 t}{D_A},
\eea
which is nothing but the late-time behaviour of Eqs.~\eref{var_eabp_reg2x}-\eref{var_eabp_reg2y}. The non-diffusive short-time regime is absent in this case.

The behaviour of the variance indicates that, in fact, in this regime, the dynamics of CRABP emulates the motion of an ABP with effective rotational diffusion constant $D_A$, which we discuss in detail in \sref{pdf_int_reg_2}. \\

\noindent{\bf Intermediate regime (III):}
In this regime, $ \dr t \gg 1 $ while $\gamma t \ll 1$. Hence, using $e^{-\dr t} \simeq 0$, irrespective of any value of $\o$, we obtain [see \aref{int_reg2} for details] an isotropic diffusive behaviour of the variance of the form
\bea
\sigma_x^2(t) = \sigma_y^2(t) \simeq 2 D_\text{CABP} t,
\label{var_reg3}
\eea
with $D_\text{CABP}$ being given by \eref{D_CABP}. 
In fact, the CRABP in this domain emulates the dynamics of a CABP that we discuss in \sref{pdf_int_reg_3}. \\

\noindent{\bf Late time regime (IV):}
In this regime, using $e^{-\dr t} \simeq 0$ and $e^{-\g t} \simeq 0$ as $\dr t \gg 1$ and $\g t \gg 1$ we find $\la x^2(t) \ra \simeq 2 \de t$ and $\la y^2(t) \ra \simeq 2 \de t$ directly from Eqs. \eqref{mnt_xa} and \eqref{mnt_ya} where $\de$ given by \eref{d_eff}. Note that we ignore the contribution of the time-independent components in both cases in the limits of $\dr t \gg 1$ and $\g t \gg 1$. In the same limits, the contribution of $\mu_x(t)$ is also negligible, and $\mu_y(t) = 0$ always. It allows us to get 
\bea
\sigma_x^2(t) = \sigma_y^2(t) \simeq 2 D_\text{eff} t
\label{var_reg4}
\eea
by using \eref{varxy}. Note that on taking $\g = 0$ and $\o =0$ in \eref{d_eff} we recover the known results of the effective diffusion constant of a CABP \cite{van2008dynamics} and an ABP \cite{basu2018active} respectively. This obtained result of $\de$ has previously been reported in Ref.\cite{haeggqwist2008hopping} and we find it consistent with the effective diffusion constant derived in Ref.\cite{weber2012active}.

It is also worth mentioning that $\de$ in \eref{d_eff} is a non-monotonic function of $\dr$. Interestingly, in the limit of $\dr \to 0$, we get $\de \simeq \g v_0^2/\o^2$, which implies a linear increase of $\de$ with the chirality reversal rate $\gamma$.
On the other hand, in the opposite limit, \ie when $\dr$ is very large one immediately finds out $\de \propto \dr^{-1}$ from  \eref{d_eff}. In an intermediate $\dr$ where $\o = (\dr + 2\g)$ we also find this $\de$ exhibits a maximum where $\de^{\text{(max)}} = v_0^2 (\dr + \g)^{-1}/4$. We illustrate this non-monotonic behaviour of $\de$ with $\dr$ in the inset of Fig.~\ref{var_crabp}(b). The solid red line in the same figure shows the variation of $\de$ for a CABP $(\g = 0)$ which is clearly smaller than that of others with non-zero $\g$ in the limit $\dr \to 0$. In the opposite limit of $\dr$, they all become $\g$ independent and collapse on each other.


Another interesting feature of $\de$ is its dependence on $\g$. With the increase of $\g$, as we mentioned, $\de$ increases for small $\dr$ [see the inset of \fref{var_crabp}(b)]. This result sharply contrasts with the result for the same of a RTP \cite{malakar2018steady} and a Direction reversing ABP (DRABP) \cite{santra2021active} where a similar dichotomous noise $\sig(t)$ is assigned that flipping it signs at the rate $\g$. Of course the flipping events in both scenarios are associated with the propulsion speed $v_0$ of the particle, unlike $\o$ in our case. In both cases of RTP $( \de = v_0^2/2 \g)$ \cite{malakar2018steady} and DRABP $( \de = v_0^2/2 (\dr + 2\g))$ \cite{santra2021active}, we observe $\de$ decreases with the increase of $\g$ which is opposite to our case of CRABP. In the opposite limit of $\dr$, \ie when $\dr \to \infty$, CRABP reduces to CABP where $\de$ becomes $\g$ independent much similar like $\de = v_0^2/2 \dr$ of DRABP that is also $\g$ independent as it reduces to an ordinary ABP \cite{basu2018active} in the same limit $\dr$.

\section{Position Distribution}
\label{pdf_pos}

The position distribution of the CRABP is defined as,
\bea
P(x,y,t) = \sum_{\sigma=\pm 1} \int d\theta \, Q_\sigma(x,y,\theta,t)
\eea
where $Q_\sigma(x,y,\theta,t)$ denotes the joint probability density corresponding to the Langevin equations \eqref{eq:model}, which satisfies the Fokker-Planck equation,
\bea
\frac {\partial Q_\sigma} {\partial t} &=& - v_0 \left (\cos \theta \frac {\partial Q_\sigma} {\partial x} + \sin \theta \frac {\partial Q_\sigma} {\partial y}\right ) - \omega \sigma \frac {\partial Q_\sigma} {\partial \theta} + \dr \frac {\partial^2 Q_\sigma} {\partial \theta^2} \cr
&& - \gamma \, Q_\sigma + \gamma \, Q_{-\sigma}.
\eea
However, it is hard to solve this Fokker-Planck equation exactly to obtain the position distribution for arbitrary values of $\gamma$ and $\dr$. Instead, we analyze the effective dynamics in the limiting scenarios where the time $t$ is well separated from both the time-scales $\dr^{-1}$ and $\gamma^{-1}$, and obtain analytical expressions for the marginal position distributions $P(x,t)$ and $P(y,t)$ using the behaviour of the active noises in these regimes. Note that, for notational simplicity, we use the same letter $P$ to denote the distribution for both $x$ and $y$, although the functional forms are different in general.

\subsection{Short-time regime (I): $t \ll \min (\g^{-1},\; \dr^{-1})$}

We start with the short-time regime $t \ll \min(\g^{-1},\; \dr^{-1})$ which corresponds to the scenario where neither the rotational Brownian motion has changed the orientation angle $\theta(t)$ appreciably, nor has the chirality reversed a significant number of times. 
To understand the behaviour of the position distribution in this regime, we first note that, from Eq.~\eqref{theta_t}, we can write,
\bea
\theta(t) = \psi(t) + \phi(t),
\eea
where, $\psi(t)$ and $\phi(t)$ denote a one-dimensional RTP and a Brownian motion, respectively,
\bea
\psi(t) &=& \omega \int_0^t ds\, \sigma(s), \cr
\phi(t) &=& \sqrt{2 D_R} \int_0^t ds\, \eta(s).
\label{psit_phit}
\eea
Then,  the active noises [see Eqs.~\eqref{eq:model} and \eqref{eff_noise}] can be recast as,
\bea
\xi_x(t) &=& v_0 [\cos \phi(t) \cos \psi(t)  - \sin \phi(t) \sin \psi(t)], \cr
\xi_y(t) &=& v_0 [\sin \phi(t) \cos \psi(t)  + \cos \phi(t) \sin \psi(t)].
\eea
In the regime, $t \ll D_R^{-1}$, the above equations can be approximated as,
\bea
\xi_x(t) &\simeq & v_0 [\cos \psi(t) - \phi(t) \sin \psi(t) ], \cr
\xi_y(t) &\simeq & v_0 [\sin \psi(t) + \phi(t) \cos \psi(t) ],
\eea
where we have used the fact that $|\phi(t)| \sim \sqrt{D_R t} \ll 1$ for  $t \ll D_R^{-1}$, and consequently,  $\cos \phi(t) \simeq 1$ and $\sin \phi(t) \simeq \phi(t)$.
Using the above results of the effective noises in the Langevin equations \eqref{eq:model}, we formally solve to get,
\bea
\label{eq:xt_short}
x(t) &=& v_0 \int_0^t ds~ \cos \psi(s) - v_0 \int_0^t ds~ \phi(s) \sin \psi(s), \\ 
y(t) &=& v_0 \int_0^t ds~ \sin \psi(s) + v_0 \int_0^t ds~ \phi(s) \cos \psi(s). \label{eq:yt_short}
\eea
From the above equations, using the Gaussian nature of the Brownian motion $\phi(s)$, one can compute the marginal distributions of the position components $x(t)$ and $y(t)$ for any given trajectory of $\psi(s).$ To this end, let us consider a trajectory of duration $[0,t]$ with $n$ chirality flips. Let $t_i$ denote the interval between the $i$-th and $(i-1)$-th flip and $\sigma_i = (-1)^{i+1} \sigma_1$ denote the sign of the chirality during this interval. Then, we have, $t = \sum_{i=1}^{n+1} t_i$ where  $t_{n+1}$ denotes the interval between the last flip and the final time $t$. It is also convenient to define $\tau_i = \sum_{j=1}^i t_i$, which denotes the time at which the $i$-th flipping event occurred; for notational consistency, we define $\tau_0=0$ and $\tau_{n+1}=t$. 
Along such a trajectory, 
\bea
\psi(s) = \omega \l( \sum_{i=1}^{j} \sigma_i t_i + \sigma_{j+1}(s - \tau_j) \r) \quad \text{for} \quad \tau_j \le s \le \tau_{j+1}. \label{eq:psi_t}
\eea
Now, from Eqs.~\eqref{eq:xt_short} and ~\eqref{eq:yt_short}, it is clear that, for such a  fixed trajectory, with a deterministic $\psi(s)$, the marginal distributions of $x(t)$ and $y(t)$ are Gaussian since both are time-integrals of the Brownian motion $\phi(s)$. The complete distribution can then be obtained by computing the contributions from all possible trajectories of $\psi(s)$. 

\begin{figure*}[t]
\centering{
\includegraphics[width=0.85\hsize]{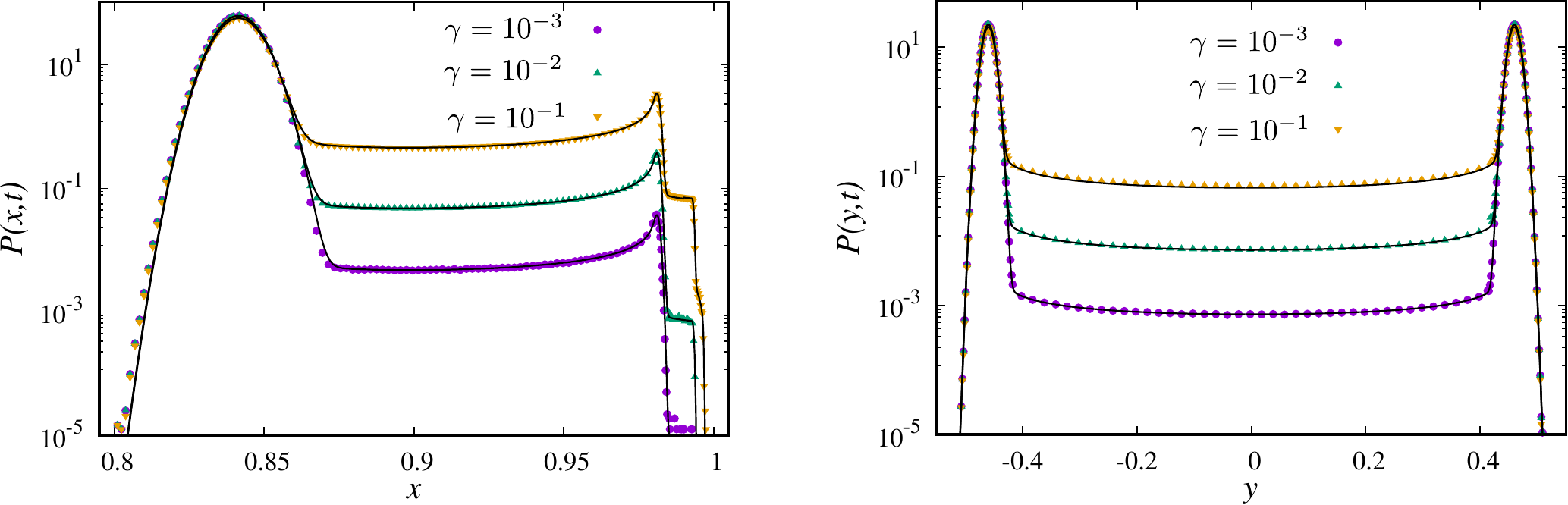}
\caption{\label{pdf_small_t} Marginal distribution of $x$ and $y$ components of position are plotted for different chirality reversing rate $\g$ at time $t = 1$ with $\dr = 0.0002$, $\o = 1$, $v_0 = 1$. Discrete symbols in each plot are simulation results which are showing excellent agreements with the analytical results [Eqs.\eqref{eq:Px_shortt} and \eqref{eq:Py_shortt} respectively] as shown by lines.} }
\end{figure*}

Let us first focus on the $x$-marginal distribution. It can formally be expressed as, 
\begin{align}
P(x,t) =& \frac 12 e^{-\gamma t} \sum_{n=0}^\infty\sum_{\sigma_1=\pm 1} \gamma^n P_{n}(x,t; \sigma_1), ~~\text{with}\cr
P_n(x,t; \sigma_1) =& \int \prod_{i=1}^{n+1}dt_i ~ \delta \l(t - \sum_{i=1}^{n+1}t_i \r) \frac {e^{-\frac{(x-x_n)^2}{2 b_n^2}}}{\sqrt{2 \pi b_n^2}}. \label{eq:Px_shortt}
\end{align}
Here, $x_n$ and $b_n^2$ are the mean and variance of the position for a given $\psi(t)$ trajectory with $n$ chirality flips, 
\bea
x_n &=& v_0 \int_0^t ds~ \cos \psi(s) = v_0 \sum_{i =1}^{n+1} \int_{\tau_{i-1}}^{\tau_i} ds\; \cos \psi(s), \cr
b_n^2 &=& v_0^2 \left \la \Big(\int_0^t ds~ \phi(s) \sin \psi(s) \Big)^2 \right \ra. 
\label{xn_bn2}
\eea
Explicit expressions for $x_n$ and $b_n^2$ can be obtained for arbitrary $n$ which are quoted in Eqs. \eqref{xnA} and \eqref{bn2A} [see \aref{pos_ini} for the details of the computation].

The formal expression for $P(x,t)$ given by Eq.~\eqref{eq:Px_shortt} is valid in the regime $t \ll D_R^{-1}$, for arbitrary $\gamma$, although the infinite sum is hard to evaluate. However, for $t \ll \gamma^{-1}$, it is expected that the typical number of chirality reversing events $n$ is small, and in the regime $t \ll (D_R^{-1}, \gamma^{-1})$ one can evaluate $P(x,t)$ explicitly using  Eq.~\eqref{eq:Px_shortt} perturbatively in $\gamma$ and truncating the sum at some small $n$. Fig.~\ref{pdf_small_t}(a) shows a plot of $P(x,t)$ in this short-time regime obtained from numerical simulations for different (small) values of $\gamma$. These are compared with the analytical prediction [Eq.~\eqref{eq:Px_shortt}] evaluated up to $n=2$ terms (details of the calculations are given in the \aref{pos_ini}); an excellent agreement validates our prediction.

It is evident from Fig.~\ref{pdf_small_t}(a) that $P(x,t)$ has a Gaussian peak at the left with a plateau around the centre of the distribution. The Gaussian peak arises from the trajectories with no reversal events up to time $t$ (\ie when $n = 0$ in \eref{eq:Px_shortt}). On the other hand, the plateau around the centre of the distribution is developed due to the trajectories with reversal events (\ie when $n \ne 0$ in \eref{eq:Px_shortt}). Particularly, for $\g = 0.1$ and $0.01$ in Fig.~\ref{pdf_small_t}(a), we clearly observe the contribution of $n = 1$ and $2$ by looking at the distribution around its left tail.

We can proceed similarly to compute the $y$-marginal distribution in this short-time regime. From Eqs. \eqref{eq:yt_short} and \eqref{eq:psi_t}, we can formally write,
\begin{align}
P(y,t) = \frac 12 e^{-\gamma t} \sum_{n=0}^\infty\sum_{\sigma_1=\pm 1} \gamma^n P_{n}(y,t; \sigma_1), \qquad \text{with,} \qquad \cr 
P_n(y,t; \sigma_1) = \int \prod_{i=1}^{n+1}dt_i ~ \delta \l(t - \sum_{i=1}^{n+1}t_i \r) \frac 1{\sqrt{2 \pi c_n^2}}\exp\left[-\frac{(y-y_n)^2}{2 c_n^2}\right],
\label{eq:Py_shortt}
\end{align}
where $y_n$ and $c_n^2$ are the mean and the variance of the $y$ component of position for a given $\psi(t)$ trajectory with $n$ chirality flips, 
\bea
y_n &=& v_0 \int_0^t ds~ \sin \psi(s) = v_0 \sum_{i =1}^{n+1} \int_{\tau_{i-1}}^{\tau_i} ds\; \sin \psi(s), \cr
c_n^2 &=& v_0^2 \left \la \Big(\int_0^t ds~ \phi(s) \cos \psi(s) \Big)^2 \right \ra. 
\eea
Following the same procedure as before, we can explicitly calculate $y_n$ and $c_n^2$ which are quoted in Eqs. \eqref{ynA} and \eqref{cn2A} in \aref{pos_ini}. Once again, we can perturbatively evaluate $P(y,t)$ in the short-time regime; see \aref{pos_ini} for the details. Fig.~\ref{pdf_small_t}(b) shows a plot of $P(y,t)$ in this short-time regime obtained from numerical simulations for different (small) values of $\g$.
Clearly, $P(y,t)$ has a double-peaked shape in this regime --- the peak positions correspond to $n = 0$ in \eref{eq:Py_shortt} which arise from the trajectories with no reversal events where the particle executes a chiral active Brownian motion. The plateau-like region around the origin is developed as chirality reversal comes into play. Particularly, this plateau-like regime around the centre comes out from the contribution of trajectories with one $(n = 1)$ or two $(n = 2)$ reversal events up to time $t$ in the limit $\g t \ll 1$.

The anisotropic nature of the chirality reversing active motion in the short-time regime (I) is clear from the qualitatively different behaviour of $P(x,t)$ and $P(y,t)$.

\begin{figure*}[t]
\centering{
\includegraphics[width=1.0\hsize]{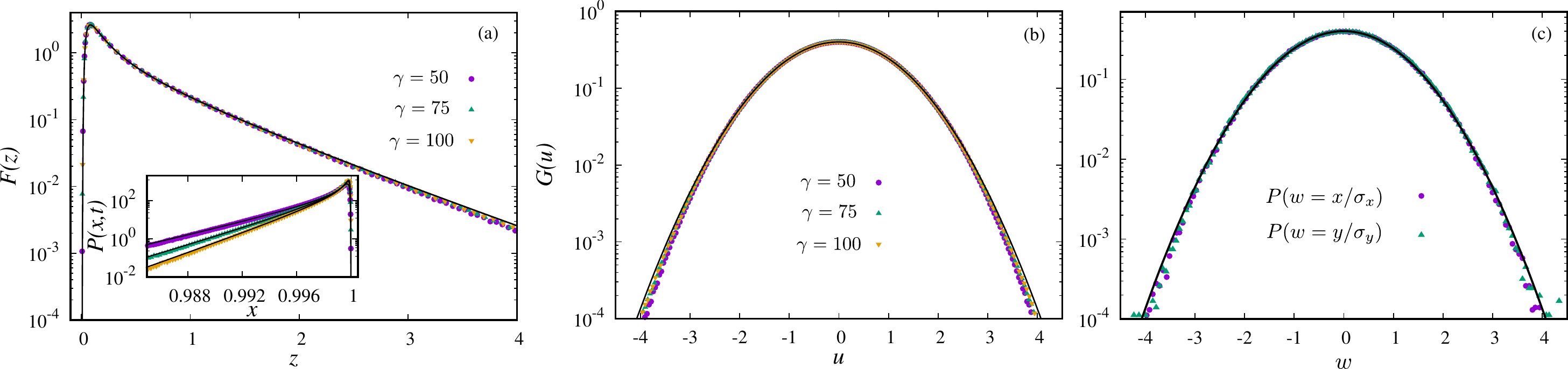}~
\caption{\label{pdf_intermediate_1} Marginal distribution of scaled $x$ and $y$ components of position in the limit $t \ll D_A^{-1}$ are plotted in (a) and (b) respectively for large values of chirality reversing rates $\g$ with $\dr = 0.001$, $\o = 0.5$ and $v_0 = 1$ at time $t = 1$ so that they satisfy both the conditions $\g^{-1} \ll t \ll \dr^{-1}$ and $\o t \sim O(1)$. Inset (a): Distribution of the unscaled $x$ component plotted for the same values of all other variables as the main plot. In (c), we plot the scaled $x$ and $y$ components of position in the limit $t \gg D_A^{-1}$ by suitably choosing $\g = 100$, $\o = 20$, $v_0 = 1$ and $\dr = 0.001$ at $t = 40$ such that $\g^{-1} \ll t \ll \dr^{-1}$ with $\o t \gg 1$. Discrete symbols are obtained from numerical simulation whereas the solid lines are the analytical results [see Eqs. \eqref{Fz}  and \eqref{pyt_2}].}} 
\end{figure*}

\subsection{Intermediate regime (II): $\g^{-1} \ll t \ll \dr^{-1}$}
\label{pdf_int_reg_2}

The intermediate regime II, which corresponds to $\g^{-1} \ll t \ll \dr^{-1}$, appears when $ \gamma \gg \dr$. For $t \gg \gamma^{-1}$, the dichotomous noise $\sigma(t)$ emulates a white noise  with zero mean and correlator $\la \sig(t) \sig(t')\ra \to \g^{-1} \delta(t - t')$ \cite{santra2021active}. Then,  for $t \gg \gamma^{-1}$, using the fact that $\eta(t)$ and $\sigma(t)$ in \eref{theta_t} are independent, the dynamics of the orientation $\theta(t)$  effectively reduces to,
\bea
\dot{\theta}(t) &=&  \sqrt{2 D_A} \; \zeta(t) ~~~~~\text{with}~~~~~ D_A =  \l(D_R + \frac{\o^2}{2\g}\r),
\label{D_A} 
\eea
where $\zeta(t)$ is a white noise with  $\la \zeta(t) \ra=0$ and $\la \zeta(t) \zeta(t') \ra = \delta(t-t')$. Eqs \eqref{vxt} and \eqref{vyt}, along with \eref{D_A}, describe the motion of an ABP with an effective rotational diffusion constant $D_A$.

Now, for $t \ll D_R^{-1}$, we also have, $D_A t \ll 1$ if $\omega t \sim O(1)$. In that case, the dynamics of the CRABP is effectively the same as that in the short-time regime $ t \ll D_A^{-1}$ of an ABP. It has been shown that the motion of an ABP is strongly anisotropic and non-diffusive in the short-time regime~\cite{basu2018active, majumdar2020toward}. In fact, from Eqs. \eqref{var_eabp_reg2x} and \eqref{var_eabp_reg2y}, we see that, $\sig_x^2(t) \propto D_A^2 t^4$ and $\sig_y^2(t) \propto D_A t^3$, which are consistent with the short-time behaviour of an ABP.


The exact analytical scaling forms for the marginal distributions in this regime have also been derived~\cite{basu2018active}. Following these results, we obtain the marginal position distributions of the CRABP in this regime. The marginal distribution of the $x$ component follows a scaling form,
\bea
P(x,t) = \frac{1}{v_0 D_A t^2}\; F\l( \frac{v_0 t - x}{v_0 D_A t^2}\r),
\eea
where the scaling function
\bea
F(z) = \frac{1}{2 \sqrt{\pi z^3}} \sum_{k=0}^{\infty} (-1)^k \frac{(4k +1)}{2^{2k}} \binom {2k}{k} \exp\l[ - \frac{(4k+1)^2}{8 z}\r],
\label{Fz}
\eea
is highly skewed, with a peak near $z = 0$. In contrast to $x$, the distribution of the $y$ component is  symmetric and Gaussian, although non-diffusive,
\bea
P(y,t) = \sqrt{\frac{3}{2 v_0^2 D_A t^3}}\; G \l( \frac{y}{v_0}\, \sqrt{\frac{3}{2D_A t^3}} \r),\cr
~\text{with}~G(u) = \frac{e^{-u^2/2}}{\sqrt{2\pi}}.
\label{pyt_2}
\eea  
We illustrate the behaviours of these two components in Fig.~\ref{pdf_intermediate_1}(a) and (b) where we have suitably chosen the parameters such that $\g t \gg 1$, $\dr t \ll 1$ as well as $t \ll D_A^{-1}$.

In the limit $t \gg D_A^{-1}$ \ie when $\o \,t \gg 1$ with $\g^{-1} \ll t \ll \dr^{-1}$, the dynamics of CRABP exhibits a  diffusive behaviour, like an ABP in the late-time regime, with an effective diffusion constant $v_0^2/(2 D_A)$. It is straightforward to find this result by using $e^{-D_A t} \simeq 0$ in Eqs.\eqref{var_eabp_reg2x} and \eqref{var_eabp_reg2y}. In this regime, the typical fluctuations along $x$ and $y$ directions are expected to be Gaussian with the variance $\sig_x^2(t) = \sig^2_y(t) \simeq v_0^2 t/D_A$. This is illustrated in Fig.~\ref{pdf_intermediate_1}(c) where we have compared the scaled distributions of $x/\sigma_x$ and $y/\sigma_y$, obtained from numerical simulations, with the standard normal distribution, which show excellent agreement.


\subsection{Intermediate regime (III): $ \dr^{-1} \ll t \ll \g^{-1}$}
\label{pdf_int_reg_3}

For $\dr \gg  \gamma$, a different scenario emerges in the intermediate regime $ \dr^{-1} \ll t \ll \g^{-1}$. 
In this regime, the effective noise components emulate white noises [see \eref{corr_reg3} in \aref{effec_noise}] with correlations,
\bea
\la \xi_x(t) \xi_x(t') \ra  = \la \xi_y(t) \xi_y(t') \ra \simeq 2 D_\text{CABP} \;\delta(t - t')
\label{tpc_reg3}
\eea 
where $D_\text{CABP}$ is given by \eref{D_CABP}. Consequently, the typical motion of the CRABP is expected to be like an ordinary Brownian particle in 2D with an effective diffusion constant $D_\text{CABP}$.

\begin{figure}[t]
\centering{
\includegraphics[width=0.8\hsize]{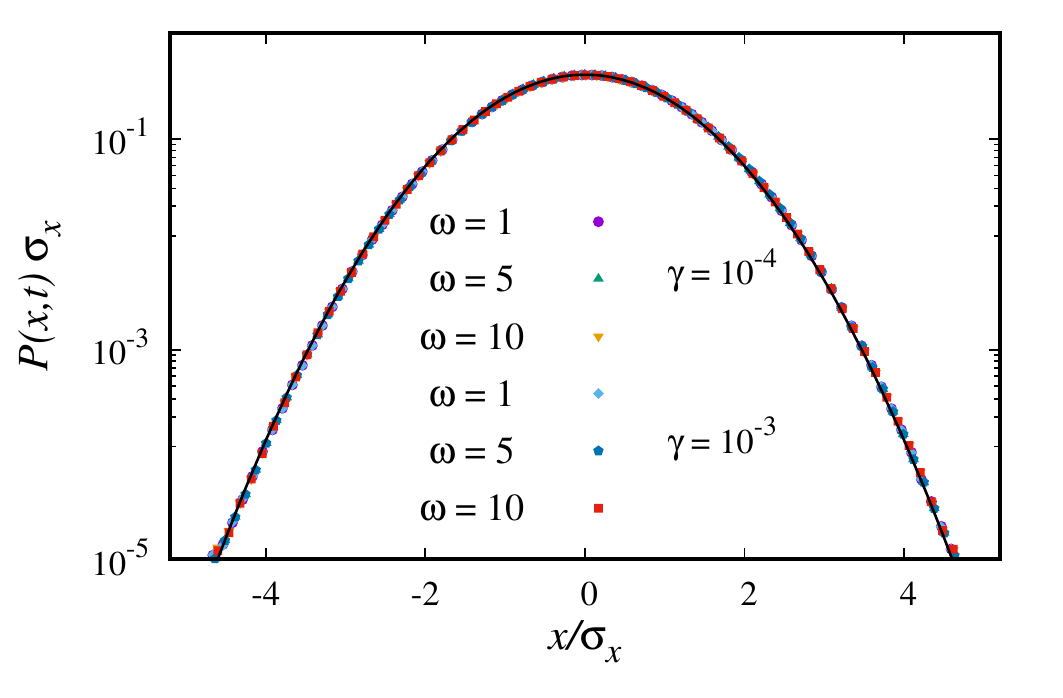}~
\caption{\label{pdf_3} Position distribution in the intermediate regime (III):  Distribution of $x$ component of position is plotted for different flipping rate $\g$ and angular speed $\o$ at time $t = 80$ with $v_0 = 1$ and $\dr = 20$. Symbols indicate the data obtained from numerical simulations while the black solid line indicates the analytically predicted Gaussian distribution [see \eref{pdf_xy_3}].} }
\end{figure}

In fact, this situation corresponds to the late-time behaviour of a chiral ABP which can be understood at follows. In this regime, the average number of reversal events $\gamma t \ll 1$, hence, to the leading order, the particle behaves like an ordinary chiral active Brownian particle. On the other hand, it is known that a CABP in the limit $t \gg \dr^{-1}$ reduces effectively to a simple Brownian motion with with the variance $\sig_x^2(t)= \sig_y^2(t) \simeq 2 D_{CABP} \;t$~\cite{van2008dynamics, ebbens2010self} which leads to the $\gamma$-independent diffusive behaviour of CRABP in this regime. Hence, the distribution of the scaled position, $x/\sig_x$ and $y/\sig_y$, becomes a standard Gaussain distribution, \ie
\begin{align}
P(x,t) &= \frac{1}{\sig_x}\; G \l(\frac{x}{\sig_x} \r)~~\text{and}~~P(y,t) = \frac{1}{\sig_y}\; G \l(\frac{y}{\sig_y} \r) \cr
 &~~~~~~~~~~~~~~~~~~~\text{with}~~G(u) = \frac{e^{-u^2/2}}{\sqrt{2\pi}}.
\label{pdf_xy_3}
\end{align}
We have plotted the distribution of the scaled $x$ component of position for differnt $\o$ and $\g$ in Fig.~\ref{pdf_3} along with the numerical simulation.

\subsection{Late-time regime (IV): $t \gg \max(\g^{-1},\; \dr^{-1})$}

\begin{figure}[t]
\centering{
\includegraphics[width=0.8\hsize]{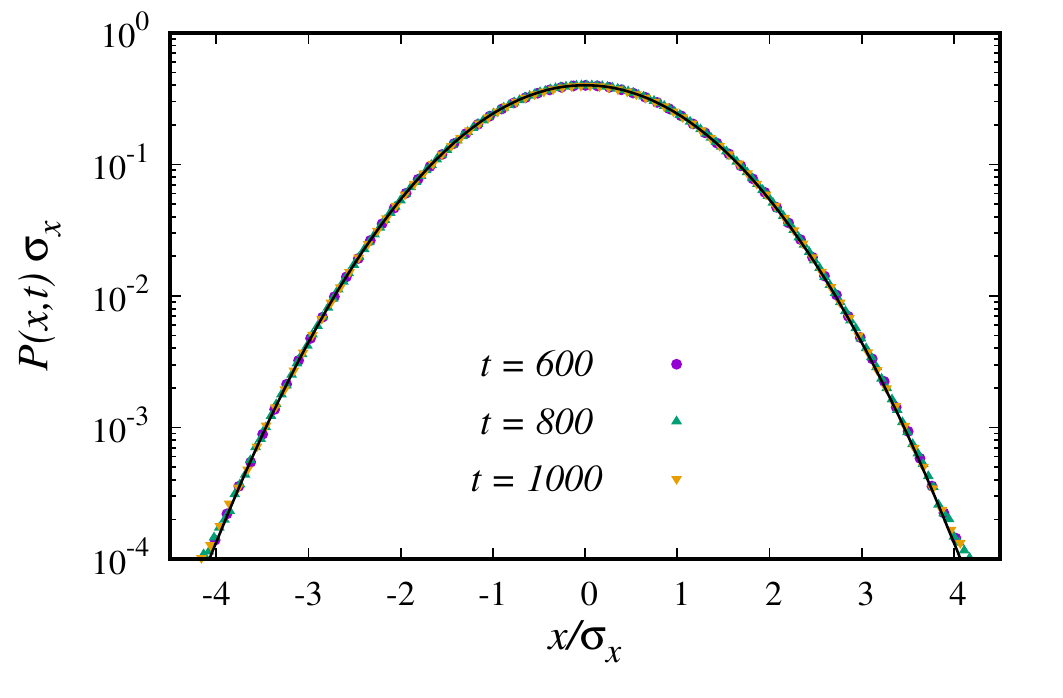}~
\caption{\label{pdf_4} Position distribution at the late time regime (IV): Distribution of $x$ component of position is plotted at different times by taking $v_0 = \o = \g = 1$ and $\dr = 10$. Discrete symbols are the simulation results which are in good agreement with the analytical result as shown by a solid line.}} 
\end{figure}

The late-time regime emerges when the time is much larger than both the time-scales of the active particle, i.e., $\g t \gg 1$ and $\dr t \gg 1$. In this regime, the two-point correlation functions of the effective noises emulate white noises [see \eref{corr_reg4} in \aref{effec_noise}],
\bea
\la \xi_x(t) \xi_x(t') \ra = \la \xi_y(t) \xi_y(t') \ra \simeq 2 \de \;\delta(t - t')
\eea 
where the strength of the noises $\de$ is nothing but the effective diffusion constant given by \eref{d_eff}. These correlations effectively reduce the dynamics of both the $x$ and $y$-components of the CRABP to simple Brownian motions. In other words, the distributions of the scaled positions, $x/\sig_x$ and $y/\sig_y$, in this regime are standard Gaussian distributions [see \eref{pdf_xy_3}]. We have illustrated this in Fig.~\ref{pdf_4} with the support of numerical simulations.

\section{Conclusions}
\label{conclusion}

In this paper, we studied the chirality reversing active motion in a two-dimensional plane using a simple model of a chiral active Brownian particle that stochastically reverses the direction of its angular velocity. The reversal of angular velocity is modeled by a dichotomous noise that flips its sign between $\pm 1$ at a constant rate $\gamma$. This reversal rate $\g$ introduces a new time-scale in addition to the one set by the rotational diffusion constant $\dr$. We find that, depending on the relative strengths of  $\dr$ and $\g$, the CRABP shows four distinct dynamical regimes: (I) $t \ll \min(\gamma^{-1}, \dr^{-1})$, (II) $ \gamma^{-1} \ll t \ll \dr^{-1}$, (III) $ \dr^{-1} \ll t \ll \gamma^{-1})$, and (IV) $t \gg \max(\gamma^{-1}, \dr^{-1})$, each characterized by different position distributions.

We find that the short-time regime is characterized by strongly anisotropic and non-diffusive position fluctuations --- while the marginal distribution along the initial orientation has a non-trivial asymmetric shape, the distribution along its orthogonal direction is symmetric with two Gaussian peaks connected by a plateau-like region across the origin. On the other hand, the typical position fluctuations of CRABP becomes diffusive and isotropic in the late-time regime, with both  $x$ and $y$-marginal distributions are Gaussian. The crossover from the short-time anisotropic to long-time isotropic distribution happens via an intermediate regime, the nature of which depends on the relative strengths of the two time-scales. For $\gamma \gg \dr$, i.e., when reversal events are frequent, the CRABP reduces to an ABP with an effective rotational diffusion constant $D_A$, which depends on $\gamma$. Depending on the angular speed $\o$, in this regime, we find an anomalous, anisotropic fluctuations of position in the limit $t \ll D_A^{-1}$, which changes to an isotropic Gaussian distribution on the opposite limit $t \gg D_A^{-1}$. The other intermediate regime appears when $\dr \gg \gamma$, where CRABP effectively becomes a CABP with an isotropic, $\gamma$-independent diffusive position fluctuations characterized by a Gaussian distribution.

Recently, the behaviour of active Brownian particles with stochastic direction reversals has been studied, and it was shown that, the presence of the additional time-scale due to direction reversal leads to a range of novel features, both in free space and in a harmonic trap~\cite{santra2021active,santra2021direction}. Here, in contrast, the active particle reverses its chirality stochastically, which leads to a very different behaviour, with different dynamical regimes. This work adds a significant contribution towards understanding the active motion in the presence of more than one intrinsic time-scales. It would be interesting to study the behaviour of CRABP in the presence of a confining potential. A related question is what happens if a CRABP is subjected to an intermittent harmonic potential, in the spirit of Ref.~\cite{santra2021brownian}. Finally, it would also be intriguing to study the collective behaviour of interacting CRABPs.


\section*{Conflicts of interest}

There are no conflicts to declare.

\section*{Acknowledgements}
U. B. acknowledges support from the Science and Engineering Re-search Board (SERB), India, under a Ramanujan Fellowship (Grant No. SB/S2/RJN-077/2018).

\appendix
\label{appendix}

\section{Properties of the effective noise}
\label{effec_noise}

In this Appendix, we explicitly compute the first two moments of the effective noise defined in Eq.~\eqref{eff_noise}. This is most conveniently done by writing the master equation for the time evolution of $P_\sigma(\t,t|\t_0, 0)$, the probability that the orientation is $\theta$ and chirality is $\sigma$ at time $t$, starting from the initial orientation $\theta_0$ at $t=0$. We assume that initial chirality can be positive or negative with equal probability, i.e., $\sigma(t=0) = \pm 1$ with probability $\frac 12$. The master equation satisfied by $P_\sigma(\t,t|\t_0, 0)$, corresponding to the Langevin equation \eqref{theta_t}, is a generalized form of telegrapher’s equation\cite{malakar2018steady},
\begin{align}
\partial_t P_\sigma(\t,t | \t_0, 0) =& - \sigma \o \, \partial_\t P_\sigma(\t,t | \t_0, 0) + \dr \,\partial_\t^2 \, P_\sigma(\t,t | \t_0, 0) \cr
& - \g \, P_\sigma(\t,t | \t_0, 0) +\g \, P_{-\sigma}(\t,t | \t_0, 0).\label{ap1}
\end{align}
To solve this coupled set of second-order partial differential equations, it is convenient to take a Fourier transformation with respect to $\theta$,
\bea
\widetilde{\mathcal{P}}_{\pm}(k,t) = \int_{-\infty}^{\infty} d\t \;e^{i k \t } P_{\pm}(\t,t | \t_0, 0).
\eea
Equation~\eqref{ap1} then reduces to a set of two coupled ordinary differential equations,
\bea
\partial_t
\left(
\begin{array}{c}
 \widetilde{\mathcal{P}}_{+}(k,t) \\
 \widetilde{\mathcal{P}}_{-}(k,t) \\
\end{array}
\right) = A
\left(
\begin{array}{c}
 \widetilde{\mathcal{P}}_{+}(k,t) \\
 \widetilde{\mathcal{P}}_{-}(k,t) \\
\end{array}
\right),
\label{matrix_1a}
\eea
where
\bea
A = 
\left(  
\begin{array}{cc}
-\dr k^2 + i k \o - \g &  \g \\
\g & -\dr k^2 - i k \o - \g\\
\end{array}
\right).
\eea 
These can be immediately solved with the boundary condition $P_{\pm}(\theta,t=0|\theta_0,0) = \delta(\theta-\theta_0)/2$ and it is convenient to express the solution as, 
\bea
\left(
\begin{array}{c}
 \widetilde{\mathcal{P}}_{+}(k,t) \\
 \widetilde{\mathcal{P}}_{-}(k,t) \\
\end{array}
\right) 
= \frac{1}{2} e^{i k \theta_0}
\left(  
\begin{array}{cc}
\widetilde{P}_{++}(k,t) \;+\; \widetilde{P}_{+-}(k,t)\\
\widetilde{P}_{-+}(k,t) \;+\;  \widetilde{P}_{--}(k,t)\\
\end{array}
\right),
\label{matrix_1a}
\eea
where
\begin{align}
\widetilde{P}_{++}(k,t) &=& e^{-(\g + \dr k^2) t} \l( \cosh[\lm_k t] + \frac{i k \o}{\lm_k} \sinh[\lm_k t] \r),\cr
\widetilde{P}_{-+}(k,t) &=& \widetilde{P}_{+-}(k,t) = e^{-(\g + \dr k^2) t} \; \frac{\g}{\lm_k} \sinh[\lm_k t],~~~~~\cr
\widetilde{P}_{--}(k,t) &=& e^{-(\g + \dr k^2) t} \l( \cosh[\lm_k t] - \frac{i k \o}{\lm_k} \sinh[\lm_k t] \r)~~\label{p_all_a}
\end{align}
with $\lm_k = \sqrt{\g^2 - k^2 \o^2}$. Note that, $\widetilde{P}_{\sig \sig_0}(k,t)$ stands for the Fourier transform of the propagator $P_{\sig \sig_0}(\t,t|\t_0 =0,0)$, which denotes the probability that the orientation is $\t$ and chirality is $\sig$ at time $t$, starting from the initial orientation $\t_0 = 0$ and chirality $\sigma_0$ at $t = 0$. The prefactor $1/2$ on the right-hand side of \eref{matrix_1a} arises from the uniform initial condition $\sig(0) = \pm 1$ with equal probability $1/2$. Finally, Fourier transform of the $\theta$-distribution can be obtained from Eqs.~\eqref{matrix_1a} and \eqref{p_all_a},
\bea
\widetilde{P}(k,t) &=& \l \la  e^{i k \t }\r \ra = \sum_{\sig} \widetilde{\mathcal{P}}_{\sig}(k,t)
= \frac{1}{2} e^{i k \t_0} \sum_{\sig, \sig_0} \widetilde{P}_{\sig \sig_0}(k,t) \cr
&=& e^{i k \t_0 - (\dr k^2 + \g)t} \left(\cosh \left[\lm_k t \right] + \frac{\gamma}{\lm_k} \sinh \left[\lm_k t \right]\right),~~~~~~~~~
\label{p_full_a}
\eea
which is nothing but the moment generating function of the $\theta$-process.
Note that the angular brackets in the above equation stand for the average over ensembles and we use this notation with the same meaning throughout this paper.

The moments and correlations of the effective noises $\xi_x(t)$ and $\xi_y(t)$ [see \eref{eff_noise}] can be computed from \eref{p_full_a} in a straightforward manner. We start with the first moments,
\begin{align}
\la \xi_x(t) \ra &=
v_0 \la \cos\t(t) \ra =  \frac{v_0}{2} \l( \widetilde{P}(1,t) + \widetilde{P}(-1,t) \r) \nonumber \\
&= v_0 \cos \t_0  \;e^{-t (\gamma + \dr)}   \left( \cosh \left( \lm_1 t \right) + \frac{\gamma \; \sinh \left(\lm_1 t \right)}{\lm_1}  \right),~~~~\label{xi_xa}
\end{align}
and,
\begin{align}
\la \xi_y(t) \ra &=
v_0 \la \sin\t(t) \ra =  \frac{v_0}{2 i} \l( \widetilde{P}(1,t) - \widetilde{P}(-1,t) \r) \nonumber \\ 
&= v_0 \sin \t_0 \;e^{-t (\gamma + \dr)} \left( \cosh \left( \lm_1 t \right) + \frac{\gamma \; \sinh \left(\lm_1 t \right)}{\lm_1}  \right).~~~~\label{xi_ya}
\end{align}
Next, we compute the two-point auto-correlations of the effective noises. For $t > t'$, the auto-correlation of $\xi_x(t)$ is given by,
\bea
 \la \xi_x(t) \xi_x(t')  \ra_{t > t'} \equiv  v_0^2 \; \la \cos\t(t) \cos \t'(t') \ra_{t > t'} ~~~~~~~~~~~~\cr
= \frac{v_0^2}{2} \; \sum_{\s, \s', \s_0 } \int_{-\infty}^{\infty} d\t  \cos \t \int_{-\infty}^{\infty}  d\t'\; P_{\s \s'}(\t,t|\t',t') \cr
\times \;\cos \t'\; P_{\s'\s_0}(\t',t'|\t_0,0). ~~~~~~~~~~~~~~~~~~~~~
\label{eq:xix_corr1}
\eea
Note that, we have considered the initial condition $\sigma_0 = \pm 1$ with equal probability $\frac 12$. Now, using $\cos z = (e^{i z} + e^{-i z})/2$,  and the definition of the following Fourier transform $P_{\s_1 \s_2}(\t,t|\t',t') = (1/2\pi)^{-1} \int_{-\infty}^{\infty} dk\; e^{-i k (\t -\t')}  \widetilde{P}_{\s_1 \s_2}(k, t - t')$, we get from \eref{eq:xix_corr1},
\begin{widetext}
\begin{align}
 \la \xi_x(t) \xi_x(t')  \ra_{t > t'} 
= & \frac{v_0^2}{8} \sum_{\s, \s', \s_0 }  \int_{-\infty}^{\infty} dk\, dk' \; \widetilde{P}_{\s \s'}(k, t - t')  \widetilde{P}_{\s'\s_0}(k',t') \; e^{i k' \t_0} \cr 
& \times \int_{-\infty}^{\infty}  \frac{d\t}{2 \pi} \; \l[ e^{- i (k-1) \t} + e^{-i  (k+1) \t}  \r]  \int_{-\infty}^{\infty}  \frac{d\t'}{2 \pi} \l[ e^{- i (k'- k - 1) \t'} + e^{-i  (k' - k + 1) \t'}  \r]. 
\end{align}
%
Clearly, the integrals over $\theta$ and $\theta'$ lead to Dirac-delta functions, which, in turn, allows trivial evaluation of the $k$ and $k'$ integrals, finally yielding,
\begin{align}
\la \xi_x(t) \xi_x(t')  \ra_{t > t'} 
= \frac{v_0^2}{8} \sum_{\s, \s', \s_0 } \sum_{k = 0}^{1} \;  \widetilde{P}_{\s \s'}(2k -1, t-t')  
\l[ \widetilde{P}_{\s' \s_0}(0,t') + \widetilde{P}_{\s' \s_0}(2(2k - 1) ,t') e^{2(2 k -1) i \t_0} \r].~~
\label{cor_xxaf}
\end{align}

The explicit form for the noise correlation can now be obtained by using \eref{p_all_a} and performing the sums over $\sigma, \sigma', \sigma_0$ and $k$, and is given by,
\bea
\la \xi_x(t) \xi_x(t') \ra_{t > t'}
= \frac{v_0^2}{2 \lm_1 \lm_2}\; e^{-\dr(t+3t') -\g t} \Bigg[ \lm_2  \l( e^{(4\dr +\g)t'}  + \cos[2 \t_0] \cosh[\lm_2 t'] \r) \l( \lm_1 \cosh[\lm_1(t - t')]  +\; \g \;\sinh[\lm_1(t - t')] \r) \cr
+ \cos[2\t_0] \sinh[\lm_2 t'] \Bigl( \g \lm_1 \cosh[\lm_1 (t-t')] + (\g^2 - 2 \o^2) \sinh[\lm_1 (t - t')] \Bigr) \Bigg], \label{cor_xxa}
\eea
where $\lm_1 = \sqrt{\g^2 - \o^2}$, and $\lm_2 = \sqrt{\g^2 - 4\o^2}$.
The auto-correlation of the $y$-component of the effective noise can also be calculated in a similar manner, and turns out to be,
\bea
\la \xi_y(t) \xi_y(t') \ra_{t > t'} 
= \frac{v_0^2}{2 \lm_1 \lm_2}\; e^{-\dr(t+3t') -\g t} \Bigg[ \lm_2  \l( e^{(4\dr +\g)t'}  - \cos[2 \t_0] \cosh[\lm_2 t'] \r) \l( \lm_1 \cosh[\lm_1(t - t')]  +\; \g \;\sinh[\lm_1(t - t')] \r) \cr 
- \cos[2\t_0] \sinh[\lm_2 t'] \Bigl( \g \lm_1 \cosh[\lm_1 (t-t')] + (\g^2 - 2 \o^2) \sinh[\lm_1 (t - t')] \Bigr) \Bigg]. \label{cor_yya}
\eea
\end{widetext}

It is important to investigate the behaviour of the effective noise in different dynamical regimes to understand the motion of CRABP in these regimes. 

To proceed, we first note that, the correlation of the effective noise becomes time translational invariant in the long-time limit $t, t' \to \infty$ while keeping $t-t'$ fixed. In this regime,  Eqs. \eqref{cor_xxa} and \eqref{cor_yya} reduce to,
\bea
\la \xi_x(t) \xi_x(t') \ra = \la \xi_y(t) \xi_y(t') \ra  
\simeq \frac{v_0^2}{4 \lm_1} \qquad \qquad \qquad \qquad \cr \times \l[ \l(\lm_1 + \gamma \r) e^{-{(\dr + \g - \lm_1) |t-t'|} } + \l(\lm_1 - \gamma \r) e^{-{(\dr + \g + \lm_1) |t-t'|}} \r]. 
\label{c4xx_c4yy}
\eea 
Now,  the long-time regime (IV) can be accessed by taking the limit $\dr \to \infty$ and $\g \to \infty$. In this regime, for arbitrary values of  $\o$, we get, from \eqref{c4xx_c4yy},
\bea
\la \xi_x(t) \xi_x(t') \ra = \la \xi_y(t) \xi_y(t') \ra 
\simeq  \frac{(\dr + 2\g) \;v_0^2}{\o^2 + \dr (\dr + 2 \g)} \delta(t -t').~~
\label{corr_reg4}
\eea
The above equation, in conjunction with central limit theorem, indicates that the effective noise emulates a white noise in regime (IV). The same statement remains valid  in regime (III), which can be accessed by taking $\gamma \to 0$ in the above equation, leading to,
\bea
\la \xi_x(t) \xi_x(t') \ra = \la \xi_y(t) \xi_y(t') \ra 
\simeq   v_0^2 \;\l( \frac{\dr}{\dr^2 + \o^2} \r) \delta(t -t').~~~
\label{corr_reg3}
\eea

\section{Computation of moments}
\label{appendix_mnts}

In this appendix, we explicitly compute  the first two moments of CRABP position components.

We start with the average position $\mu_x(t)$ and $\mu_y(t)$, which can be computed in a straightforward manner using Eqs. \eqref{xi_xa} and \eqref{xi_ya},
\begin{align} 
\mu_x(t) = \int_0^t d\tau\; \la \xi_x(t) \ra =\frac{v_0 (2 \g + \dr) \cos \t_0}{\l(\dr^2 + 2 \g \dr + \o ^2\r)} \l[  \l( 1 - e^{-t (\g + \dr)} \r. \r.~~ \cr
\l. \l. \times \cosh \l( \lm_1 t \r) \r) 
+ \l( \frac{\o^2}{(2 \g + \dr)} -\g \r) e^{-t (\g + \dr)} \frac{\sinh \l( \lm_1 t \r)}{\lm_1} \r],
\label{mu_xta}
\end{align}
and
\begin{align}
\n
\mu_y(t) = \int_0^t d\tau\; \la \xi_y(t) \ra = \frac{v_0 (2 \g + \dr) \sin \t_0}{\l(\dr^2 + 2 \g \dr + \o ^2\r)} \l[  \l( 1 - e^{-t (\g + \dr)} \r. \r.~~ \\ \l. \l. \times \cosh \l( \lm_1 t \r) \r) 
+ \l( \frac{\o^2}{(2 \g + \dr)} -\g \r) e^{-t (\g + \dr)} \frac{\sinh \l( \lm_1 t \r)}{\lm_1} \r].
\label{mu_yta}
\end{align}
For $\theta_0=0$, $\mu_y(t)$  remains zero at all times, and $\mu_x(t)$ reduces to a simpler expression, which is quoted  in Eq.~\eqref{mu_xt}.

The second moments of the position components are given by,
\bea
\label{2nd_mntx}
\la x^2(t) \ra &=& 2 \int_0^t d\tau \int_0^{\tau} d\tau' \;
\la \xi_x(\tau) \xi_x(\tau') \ra_{\tau > \tau'},\\
\la y^2(t) \ra &=& 2 \int_0^t d\tau \int_0^{\tau} d\tau' \;
\la \xi_y(\tau) \xi_y(\tau') \ra_{\tau > \tau'}. 
\label{2nd_mnty}
\eea
Now using $\la \xi_x(\tau) \xi_x(\tau') \ra_{\tau > \tau'}$ and $\la \xi_y(\tau) \xi_y(\tau') \ra_{\tau > \tau'}$ from Eqs. \eqref{cor_xxa} and \eqref{cor_yya} with $\t_0 = 0$ we obtain
\bea
\n
\la x^2(t) \ra &=& v_0^2 \l( \frac{\dr + 2\g}{\o^2 + \dr( 2 \g + \dr)} \r) \;t - v_0^2 \l [ A_0 - 2 A_1\r] \\ \n
&+& v_0^2  \l [ A_0 +\frac{1}{2\lm_1} (A_2 - A_3) \r] e^{-(\dr+\g)t} \cosh(\lm_1 t) \\ \n
&+& v_0^2  \l [ A_4 -\frac{1}{2\lm_1} (A_2 + A_3) \r] e^{-(\dr+\g)t} \sinh(\lm_1 t) \\ \n
&+& \frac{v_0^2}{2 \lm_2}  \l (A_5 + A_6 \r) e^{-(4\dr+\g)t} \cosh(\lm_2 t) \\
&-& \frac{v_0^2}{2 \lm_2}  \l (A_5 - A_6 \r) e^{-(4\dr+\g)t} \sinh(\lm_2 t), 
\label{mnt_xa}
\eea
and
\bea
\n
\la y^2(t) \ra &=& v_0^2 \l( \frac{\dr + 2\g}{\o^2 + \dr( 2 \g + \dr)} \r) \;t - v_0^2 \l [ A_0 + 2 A_1\r] \\ \n
&+& v_0^2  \l [ A_0 -\frac{1}{2\lm_1} (A_2 - A_3) \r] e^{-(\dr+\g)t} \cosh(\lm_1 t) \\ \n
&+& v_0^2  \l [ A_4 +\frac{1}{2\lm_1} (A_2 + A_3) \r] e^{-(\dr+\g)t} \sinh(\lm_1 t) \\ \n
&-& \frac{v_0^2}{2 \lm_2}  \l (A_5 + A_6 \r) e^{-(4\dr+\g)t} \cosh(\lm_2 t) \\
&+& \frac{v_0^2}{2 \lm_2}  \l (A_5 - A_6 \r) e^{-(4\dr+\g)t} \sinh(\lm_2 t) 
\label{mnt_ya}
\eea
where $\lm_1 = \sqrt{\g^2 - \o^2}$, $\lm_2 = \sqrt{\g^2 - 4\o^2}$ and $\de$ is given by \eref{d_eff} in the main text with coefficients
\begin{subequations}
\bea
\label{A0}
A_0 &=& \frac{(\dr + \g)(\dr + 3 \g)+ \lm_1^2}{((\dr + \g)^2 - \lm_1^2)^2},\\
\label{A1}
A_1 &=& \frac{(\dr + \g)(4 \dr + \g) +\lm_1^2 - 2\dr^2}{((\dr + \g)^2 - \lm_1^2)((4\dr + \g)^2 - \lm_2^2)},\\
A_2 &=& \frac{(\g - \lm_1) (3\dr - 3 \lm_1 -\g)}{(\dr+\g+\lm_1) ((3\dr-\lm_1)^2 - \lm_2^2)},\\
A_3 &=& \frac{(\g + \lm_1) (3\dr + 3 \lm_1 -\g)}{(\dr+\g-\lm_1) ((3\dr+\lm_1)^2 - \lm_2^2)},\\
\label{A4}
A_4 &=& \frac{\g(\dr + \g)^2 + (2\dr + 3 \g) \lm_1^2}{\lm_1 ((\dr + \g)^2 - \lm_1^2)^2},\\
A_5 &=& \frac{3\dr(\lm_2-\g)+\lm_2^2 +\g^2 - 2 \lm_2 \g - 2 \lm_1^2}{(4\dr+\g+\lm_2) ((3\dr+\lm_2)^2 - \lm_1^2)},\\
A_6 &=& \frac{3\dr(\lm_2+\g)-\lm_2^2 +\g^2 - 2 \lm_2 \g - 2 \lm_1^2}{(4\dr+\g-\lm_2) ((3\dr-\lm_2)^2 - \lm_1^2)}.
\eea
\label{AA}
\end{subequations}
Interestingly, the second moment of the radial distance of the particle $\la r^2 \ra = \la x^2 \ra + \la y^2 \ra$ reduces to a simple form,
\bea
\n
\la r^2(t) \ra = \frac{2 v_0^2 (\dr + 2\g)\; t}{\o^2 + \dr( 2 \g + \dr)} + 2 A_4 v_0^2 \;e^{-(\dr+\g)t} \sinh(\lm_1 t) \\   
- 2 A_0 v_0^2 \l( 1 - e^{-(\dr+\g)t} \cosh(\lm_1 t)\r).
\eea
Using $A_0$ and $A_4$ from Eqs. \eqref{A0} and \eqref{A4} we further get an explicit form of $\la r^2(t) \ra$ which is quoted in \eref{mnt2_r}.

In the following we investigate the behaviour of the position moments in the two intermediate regimes.

\subsection{Variance in intermediate regime (II)}
\label{int_reg1}

This intermediate regime (II), where $\g^{-1} \ll t \ll \dr^{-1}$, appears when  $\g \gg \dr$ or $\alpha = \dr/\g \ll 1$. Now, two different scenarios emerge depending on whether $\o$ is larger or smaller than $\gamma$. Let us first consider the scenario $\g > \o$ (or $\nu = \o/\g < 1$). In this case, from \eref{mu_xt} we get,
\bea
\mu_x(t) = \frac{v_0}{D_A} \l( 1 - e^{-D_A t}\r) + O(\alpha, \nu^4),
\label{muxt_muyt_a2}
\eea
where $D_A = \dr + \o^2/2\g$. Moreover, to compute the behaviour of the second moment, we note that, in this regime, $e^{-\g t} \simeq 0$, as $\g t \gg 1$, and to the leading order in $\nu$,  $\lambda_1 \simeq \g (1 - \nu^2/2)$ and $\lambda_2 \simeq \g (1 - 2 \nu^2)$. Using these in, Eqs.~\eqref{mnt_xa} and \eqref{mnt_ya}, we get,
\begin{subequations}
\bea
\n
\la x^2 (t)\ra = \frac{v_0^2 t}{D_A} - v_0^2 (A_0 - 2 A_1) + \frac{v_0^2 \; A_6}{ 2 \g} (1 + 2 \nu^2)\;  e^{-4 D_A t} \\
+ \frac{1}{2} v_0^2 \l( A_0 + A_4 - \g^{-1} (1 + \nu^2/2)\; A_3 \r)\; e^{-D_A t} + O(\alpha, \nu^4),\\
\n
\la y^2 (t)\ra = \frac{v_0^2 t}{D_A} - v_0^2 (A_0 + 2 A_1) - \frac{v_0^2 \; A_6}{ 2 \g} (1+ 2 \nu^2)\; e^{-4 D_A t} \\
+ \frac{1}{2} v_0^2 \l( A_0 + A_4 + \g^{-1} (1 + \nu^2/2)\; A_3 \r)\; e^{-D_A t} + O(\alpha, \nu^4).
\eea
\label{x2t_y2t_a2}
\end{subequations}
Taking both $\alpha \to 0$ and $\nu \to 0$  in \eref{AA} we get 
\begin{subequations}
\bea
A_0 = 8 A_1 = A_4 &=&  \frac{1}{ D_A^2} + O(\alpha, \nu^2),\\
A_3 = 4 A_6 &=& \frac{2 \g}{3 D_A^2} + O(\alpha, \nu^2).
\eea
\end{subequations}
Using them in \eref{x2t_y2t_a2} we further compute the second moments of position of $x$ and $y$ components as 
\begin{align}
\label{x2t_y2t_a2f1}
\la x^2 (t)\ra = \frac{v_0^2 t}{D_A} + \frac{v_0^2}{12 D_A^2}\;\l( e^{-4 D_A t} + 8 \;e^{- D_A t} - 9 \r)+ O(\alpha, \nu^2),~~~\\
\la y^2 (t)\ra = \frac{v_0^2 t}{D_A} - \frac{v_0^2}{12 D_A^2}\;\l( e^{-4 D_A t} - 16 \;e^{- D_A t} + 15\r) + O(\alpha, \nu^2).
\label{x2t_y2t_a2f2}
\end{align}
Using Eqs.~\eqref{varxy} and \eqref{muxt_muyt_a2} along with Eqs. \eqref{x2t_y2t_a2f1} and \eqref{x2t_y2t_a2f2}, we obtain the variance in this regime which is quoted in Eqs.~\eqref{var_eabp_reg2x} and \eqref{var_eabp_reg2y} in \sref{mean_var_pos}. On the other hand, the variance shows a linear growth in time for $\nu = \o/\gamma > 1$, that we discuss in the main text.



\subsection{Variance in intermediate regime (III)}
\label{int_reg2}

In the intermediate regime (III) where $ \dr^{-1} \ll t \ll \g^{-1}$, using $\dr t \gg 1$, that corresponds $e^{-\dr t} \simeq 0$, with  a positive $\chi = \g/\dr \ll 1$ as $\dr \gg \g$, we directly compute 
\begin{align}
A_0 \simeq \frac{\dr^2 - \o^2}{\left(\dr^2 + \o^2 \right)^2} - \frac{8 \dr^4 \chi}{\left(\dr^2 + \o^2\right)^3},~~~~~~~~~~~~~~~~~~~~\\
A_1 \simeq \frac{2 \dr^2 - \o^2}{4 \left(\dr^2+ \o^2\right) \left(4 \dr^2 + \o^2\right)} + \frac{9 \dr^2 \o^2 \left(3 \dr^2 + \o^2\right) \chi }{4 \left(\dr^2 + \o^2\right)^2 \left(4 \dr^2 + \o^2\right)^2},
\end{align}
from Eqs. \eqref{A0} and \eqref{A1} respectively. In both cases, irrespective of the value of $\o$, the first term on the right-hand side dominates over the second term as $\chi \ll 1$. Employing the same limits in Eqs. \eqref{mnt_xa} and \eqref{mnt_ya} we get 
\begin{align}
\la x^2(t) \ra \simeq v_0^2 \l( \frac{ \;\dr}{\dr^2 + \o^2} + \frac{2 \dr \o^2 z}{(\dr^2 + \o^2)^2}  \r)  \;t - v_0^2 \l[ A_0 - 2 A_1\r],\\
\la y^2(t) \ra \simeq v_0^2 \l( \frac{ \;\dr}{\dr^2 + \o^2} + \frac{2 \dr \o^2 z}{(\dr^2 + \o^2)^2}  \r)  \;t - v_0^2 \l[ A_0 + 2 A_1\r].
\label{axyt_reg3}
\end{align}
As $\chi \ll 1$, using Eqs.\eqref{mu_xt}, \eqref{mu_yt}, \eqref{varxy} alongside the above results, we obtain 
\begin{align}
\sigma_x^2(t) \simeq \frac{ v_0^2 \;\dr}{\dr^2 + \o^2}\;t 
+ v_0^2 \l( \frac{3 \o^2 + \dr^2}{2 (\dr^2 + \o^2)^2} - \frac{1}{ 4 \dr^2 + \o^2} \r) + O(\chi),\\
\sigma_y^2(t) \simeq \frac{ v_0^2 \;\dr}{\dr^2 + \o^2}\;t
+ v_0^2 \l( \frac{\o^2 -3 \dr^2}{2 (\dr^2 + \o^2)^2} + \frac{1}{ 4 \dr^2 + \o^2} \r)  + O(\chi).
\end{align}
For $\dr t \gg 1$, we get a diffusive behaviour of the variance of the form $\sigma_x^2(t) = \sigma_y^2(t) \simeq 2 D_\text{CABP} t$ with $D_\text{CABP}$ given by \eref{D_CABP}.

\section{Position distribution in the short-time regime}
\label{pos_ini}

In this Appendix we provide details of the perturbative computation of the short-time position distribution. 
The mean displacement of the $x$ component along a trajectory with $n$ chirality reversals is given by \eref{xn_bn2},
\bea
x_n= v_0 \int_0^t ds~ \cos \psi(s) = v_0 \sum_{i =1}^{n+1} \int_{\tau_{i-1}}^{\tau_i} ds\; \cos \psi(s),
\eea
Using Eq.~\eqref{eq:psi_t} we get,
\begin{align}
x_n &= v_0 \sum_{i=1}^{n+1} \int_{\tau_{i-1}}^{\tau_i} \cos [\omega ( g_{i-1} + \sigma_i (s- \tau_{i-1}))] \cr
&= \frac{v_0}{\omega} \sum_{i=1}^{n+1} \sigma_i \Big[-\sin (\omega g_{i-1}) + \sin (\omega g_i) \Big] \label{eq:xnA}
\end{align}
where we have introduced the notation,
\bea
g_i = \sum_{j=1}^{i} \sigma_j t_j 
\eea 
and used the fact that $\sigma_i^{-1} = \sigma_i$. Remembering that $\sigma_{i+1} = -\sigma_i$, Eq.~\eqref{eq:xnA} can be further simplified to get,
\bea
x_n = \left \{
\begin{split}
&\frac{v_0}{\omega} \sin (\omega t)  \qquad \text{for} ~ n=0 \cr
&\frac{v_0}{\omega} \left [2 \sum_{i=1}^n \sigma_i \sin (\omega g_i) + \sigma_{n+1} \sin (\omega g_{n+1}) \right] ~ \text{for}~ n>0.
\end{split}
\right.
\label{xnA}
\eea
Next, we proceed to compute the second moment $b_n^2$ of the $x$-component of the position. From \eref{xn_bn2},
\begin{align}
b_n^2 &= v_0^2  \left \la  \left [\sum_{i=1}^{n+1} \int_{\tau_{i-1}}^{\tau_i} ds~ \phi(s) \sin \psi(s)  \right ]^2 \right \ra \cr
&= v_0^2 \sum_{i=1}^{n+1} \sum_{j=1}^{n+1} \int_{\tau_{i-1}}^{\tau_i} ds \int_{\tau_{j-1}}^{\tau_j}  ds' ~ \la \phi(s) \phi(s') \ra \cr
\times&  \sin[\omega (g_{i-1} + \sigma_i (s-\tau_{i-1}))] \sin[\omega (g_{j-1} + \sigma_j (s'-\tau_{j-1}))].~~~~ \label{eq:bn1}
\end{align}
It is convenient to separate the $i\ne j$ and $i=j$ terms of the double sum in the above equation. Moreover, since $\phi(s)$ is an ordinary Brownian motion, $\la \phi(s) \phi(s') \ra = 2 D_R \min(s,s')$. Then, using the fact that the summand is symmetric in $(i,j)$, we have, from Eq.~\eqref{eq:bn1},
\begin{align}
\n
b_n^2 &= 4 v_0^2 D_R \Bigg [\sum_{i=2}^{n+1}\sum_{j=1}^{i-1}\int_{\tau_{i-1}}^{\tau_i} ds~ \sin[\omega (g_{i-1} + \sigma_i (s-\tau_{i-1}))] \\ \n
& \times~  \int_{\tau_{j-1}}^{\tau_j}  ds' s'  ~\sin[\omega (g_{j-1} + \sigma_j (s'-\tau_{j-1}))]  \\ \n
& +~  \sum_{i=1}^{n+1}  \int_{\tau_{i-1}}^{\tau_i} ds ~\sin[\omega (g_{i-1} + \sigma_i (s-\tau_{i-1}))] \\
& \times~  7  \int_{\tau_{i-1}}^{s}  ds' s'  \sin[\omega (g_{i-1} + \sigma_i (s'-\tau_{i-1}))] \Bigg ]. 
\end{align}
The integrals can be computed exactly, and a few steps of algebra lead to,
\begin{align}
b_n^2 &= \frac{v_0^2 D_R}{2 \omega^3}\l [16 \omega \sum_{i=3}^{n+1} \sum_{j=1}^{i-2} \sigma_i \sigma_j \tau_j \cos \omega g_j  (\cos \omega g_i - \cos \omega g_{i-1})  \r. \n \\ 
&\l. \n +~~  4 \omega t  - 3 \sigma_{n+1} \sin (2 \omega g_{n+1}) - 6 \sum_{i=1}^n \sigma_i \sin (2 \omega g_i)  \r. \\ & \l. +~~ 2 \omega t \cos (2\omega g_{n+1})  + 16 \omega \sum_{i=1}^n \tau_i \cos \omega g_i (\cos \omega g_i - \cos \omega g_{i+1})  \r].
\label{bn2A}
\end{align}

Explicitly, for $n=0$, 
\bea
b_0^2 &=& \frac {v_0^2 D_R}{2 \omega^3} \Big[ 2 \omega t (2 + \cos (2 \omega t))-3 \sin (2 \omega t) \Big]
\eea
with $x_0$ given in \eref{xnA}.
For $n = 1$ we get 
\begin{align}
\n
x_1 &=  \frac{v_0}{ \omega} \left(\sin ( \o (t-2 t_1))+2\sin ( \o t_1)\right), \\ \n
b^2_1 &= \frac {v_0^2 D_R}{2  \omega^3} \l[ 4 \omega (t+2 t_1)-3 \sin (2 \omega (t-2 t_1)) \r. \\  &\l.\n -~ 8 \omega t_1  \{ \cos (\o (t-t_1))+\cos (\omega (t-3 t_1))-\cos (2 \omega t_1 ) \} \r. \\ & \l. +~ 2 \omega t \cos (2 \omega (t-2 t_1))-6 \sin (2 \omega t_1 ) \r]
\end{align}
with,
\bea
P_1(x,t) = \int_0^t dt_1~ \frac 1{\sqrt{2 \pi b_1^2}}\exp{- \frac{(x-x_1)^2}{2 b_1^2}}.
\eea

For $n=2$, i.e., two chirality flipping events at time $\tau_1 =t_1$ and $\tau_2 = t_1 + t_2$, we have from \eref{eq:Px_shortt},
\bea
P_2(x,t) = \int_0^t dt_1 \int_0^{t-t_1} dt_2  \frac 1{\sqrt{2 \pi b_2^2}}\exp{- \frac{(x-x_2)^2}{2 b_2^2}},
\eea
with
\bea
x_2 =  \frac{v_0}{ \omega} \l[ \sin (\o (t- 2 t_2)) + 2 \sin (\o (t_2 -t_1)) + 2 \sin (\o t_1 )\r],
\eea
and
\begin{align}
\n
b^2_2 &= \frac {v_0^2 D_R}{2  \omega^3} \l[ 4 \o (t + 4 t_1 + 2 t_2) -6 \sin (2 \o t_1 ) + 6 \sin (2 \o (t_1 - t_2)) ~ \r. \\ &\l.\n -~~ 3 \sin (2 \o (t - 2 t_2)) + 2 \o t \cos (2 \o (t - 2 t_2)) + 8 \o \{ t_1 \cos (2 \o t_1 ) \r. \\ &\l. \n +~~ t_1 \cos (\o (t + t_1 - 2 t_2)) -t_1 \cos (\o (t+t_1-3t_2)) \r. \\ &\l.\n -~~ t_2 \cos (\o (t + t_1 - 3 t_2)) - (t_1 + t_2) \cos (\o (t- t_1 - t_2)) \r. \\ & \l.\n +~~ t_1 \cos (2 \o (t_1 - t_2)) + t_2 \cos (2 \o (t_1- t_2)) - 2 t_1 \cos (\o t_2 ) \r. \\ & \l. +~~ t_1 \cos (\o (-t+ t_1 + 2 t_2)) -2 t_1 \cos (\o (2 t_1 -t_2))  \} \r].
\end{align}

Note that, for our choice of initial condition, both $x_n$ and $b_n^2$ are invariant under $\sigma_1 \to  -\sigma_1$, and hence independent of $\sigma_1$. The same will not be true for $y$. For the $y$ component, for an arbitrary $n$ we simplify
\bea
y_n = \frac{ v_0}{\o} \sum_{i=1}^{n+1} \sig_i \left [ \cos (\omega g_{i-1}) -  \cos (\omega g_{i}) \right].
\label{eq:ynA}
\eea
It can further be simplified to get,
\bea
y_n = \left \{
\begin{split}
&\frac{v_0}{\omega} \sig_1 \l[ 1 - \cos (\omega t)\r]  \qquad \text{for} ~ n=0 \\
&\frac{v_0}{\omega} \left [ \sig_1 - 2 \sum_{i=1}^n \sigma_i \cos (\omega g_i) - \sigma_{n+1} \cos (\omega g_{n+1}) \right] \\&~~~~~~~~~~~~~~~~~~~~~~~~~~~~~~~~~~~~ ~ \text{for}~ n >0.
\end{split}
\right.
\label{ynA}
\eea

Following a similar method of how we computed $b_n^2$, we find
\begin{align}
c_n^2 &= 4 v_0^2 D_R \Bigg [\sum_{i=2}^{n+1}\sum_{j=1}^{i-1}\int_{\tau_{i-1}}^{\tau_i} ds~ \cos[\omega (g_{i-1} + \sigma_i (s-\tau_{i-1}))] \cr
& \times ~~ \int_{\tau_{j-1}}^{\tau_j}  ds' s'  ~\cos[\omega (g_{j-1} + \sigma_j (s'-\tau_{j-1}))]  \cr
& + ~~\sum_{i=1}^{n+1}  \int_{\tau_{i-1}}^{\tau_i} ds ~\cos[\omega (g_{i-1} + \sigma_i (s-\tau_{i-1}))] \cr
&  \times ~~ \int_{\tau_{i-1}}^{s}  ds' s'  \cos[\omega (g_{i-1} + \sigma_i (s'-\tau_{i-1}))] \Bigg ]. 
\end{align}
A few steps of algebra from here give us
\begin{align}
\n
c_n^2 &= \frac{v_0^2 D_R}{2 \omega^3}\l [16 \omega \sum_{i=3}^{n+1} \sum_{j=1}^{i-2} \sigma_i \sigma_j \tau_j \sin \omega g_j  (\sin \omega g_i - \sin \omega g_{i-1}) \r. \\ &\l. \n +~~ 4 \omega t + 3 \sigma_{n+1} \sin (2 \omega g_{n+1}) + 6 \sum_{i=1}^n \sigma_i \sin (2 \omega g_i)  \r. \\ &\l.\n -~~ 2 \omega t \cos (2\omega g_{n+1}) + 16 \omega \sum_{i=1}^n \tau_i \sin \omega g_i (\sin \omega g_i - \sin \omega g_{i+1}) \r. \\ &\l. 
-~~ 16 \sum_{i=1}^n  \sig_i \sin \o g_i - 8 \sig_{n+1} \sin \o g_{n+1} \r].
\label{cn2A}
\end{align}

For $n=0$, we find 
\bea
c_0^2 = \frac {v_0^2 D_R}{2 \omega^3} \Big[ 2 \omega t (2 - \cos (2 \omega t)) + 3 \sin (2 \omega t) - 8 \sin(\o t) \Big]
\eea
with $y_0$ given in \eref{ynA}. Similarly, for $n = 1$ we get
\begin{align}
\n
y_1 &=  \frac{v_0}{ \o \sig_1} \left(1 + \cos ( \o (t-2 t_1))- 2 \cos ( \o t_1)\right),~~ \\ \n
c^2_1 &= \frac {v_0^2 D_R}{2  \omega^3} \l[ 4 \omega (t+2 t_1) + 3 \sin (2 \omega (t-2 t_1)) ~~~ \r. \\ & \l.\n -~~ 8 \omega t_1  \{\cos (\o (t-t_1)) - \cos (\omega (t-3 t_1)) + \cos (2 \omega t_1 ) \} \r. \\ & \l. -~~ 2 \omega t \cos (2 \omega (t-2 t_1)) + 6 \sin (2 \omega t_1 ) - 16 \sin (\omega t_1 ) \r],~~
\end{align}
with
\bea
P_1(y,t) = \int_0^t dt_1~ \frac 1{\sqrt{2 \pi c_1^2}}\exp{- \frac{(y-y_1)^2}{2 c_1^2}}
\eea
from \eref{eq:Py_shortt}.

Finally, for $n = 2$, i.e., when two chirality flipping events occur at time $\tau_1 =t_1$ and $\tau_2 = t_1 + t_2$, we have from \eref{eq:Py_shortt},
\bea
P_2(y,t) = \int_0^t dt_1 \int_0^{t-t_1} dt_2  \frac 1{\sqrt{2 \pi c_2^2}}\exp{- \frac{(y-y_2)^2}{2 c_2^2}},
\eea
with
\begin{align}
y_2 =  \frac{v_0}{ \omega} \sig_1 \l[1 -\cos (\o (t - 2 t_2)) + 2 \cos (\o (t_2 - t_1)) - 2 \cos (\o t_1 )\r],
\end{align}
and
\begin{align}
\n
c^2_2 &= \frac {v_0^2 D_R}{2  \omega^3} \l[ 4 \o (t + 4 t_1 + 2 t_2) + 6 \sin (2 \o t_1 ) - 16 \sin (\o t_1) \r. \\ &\l.\n -~~ 8 \sin (\o (t - 2 t_2)) + 3 \sin (2 \o (t - 2 t_2)) + 16 \sin (\o (t_1 -t_2))  \r. \\ &\l.\n -~~ 6 \sin (2 \o (t_1 - t_2)) - 2 \o t \cos (2 \o (t-2 t_2)) - 8 \o \{ t_1 \cos (2 \o t_1 ) \r. \\ &\l.\n -~~ t_1 \cos (\o (t + t_1 -3 t_2))- t_2 \cos (\o (t+ t_1 - 3 t_2)) \r. \\ &\l.\n +~~ t_1 \cos (\o (t + t_1 - 2 t_2))- t_1 \cos (\o (t- t_1 - 2 t_2)) \r. \\ &\l.\n +~~(t_1 + t_2) \cos (\o (t-t_1- t_2))+ t_1 \cos (2 \o (t_1- t_2)) \r. \\ &\l. +~~ t_2 \cos (2 \o (t_1- t_2)) - 2 t_1 \cos (\o (2 t_1-t_2)) + 2 t_1 \cos (\o t_2 ) \} \r].
\end{align}


\bibliography{crabp_bib}

\providecommand{\noopsort}[1]{}\providecommand{\singleletter}[1]{#1}%
\providecommand*{\mcitethebibliography}{\thebibliography}
\csname @ifundefined\endcsname{endmcitethebibliography}
{\let\endmcitethebibliography\endthebibliography}{}
\begin{mcitethebibliography}{38}
\providecommand*{\natexlab}[1]{#1}
\providecommand*{\mciteSetBstSublistMode}[1]{}
\providecommand*{\mciteSetBstMaxWidthForm}[2]{}
\providecommand*{\mciteBstWouldAddEndPuncttrue}
  {\def\EndOfBibitem{\unskip.}}
\providecommand*{\mciteBstWouldAddEndPunctfalse}
  {\let\EndOfBibitem\relax}
\providecommand*{\mciteSetBstMidEndSepPunct}[3]{}
\providecommand*{\mciteSetBstSublistLabelBeginEnd}[3]{}
\providecommand*{\EndOfBibitem}{}
\mciteSetBstSublistMode{f}
\mciteSetBstMaxWidthForm{subitem}
{(\emph{\alph{mcitesubitemcount}})}
\mciteSetBstSublistLabelBeginEnd{\mcitemaxwidthsubitemform\space}
{\relax}{\relax}

\bibitem[Bechinger \emph{et~al.}(2016)Bechinger, Di~Leonardo, L{\"o}wen,
  Reichhardt, Volpe, and Volpe]{bechinger2016active}
C.~Bechinger, R.~Di~Leonardo, H.~L{\"o}wen, C.~Reichhardt, G.~Volpe and
  G.~Volpe, \emph{Rev. Mod. Phys.}, 2016, \textbf{88}, 045006\relax
\mciteBstWouldAddEndPuncttrue
\mciteSetBstMidEndSepPunct{\mcitedefaultmidpunct}
{\mcitedefaultendpunct}{\mcitedefaultseppunct}\relax
\EndOfBibitem
\bibitem[Marchetti \emph{et~al.}(2013)Marchetti, Joanny, Ramaswamy, Liverpool,
  Prost, Rao, and Simha]{marchetti2013hydrodynamics}
M.~C. Marchetti, J.-F. Joanny, S.~Ramaswamy, T.~B. Liverpool, J.~Prost, M.~Rao
  and R.~A. Simha, \emph{Rev. Mod. Phys.}, 2013, \textbf{85}, 1143\relax
\mciteBstWouldAddEndPuncttrue
\mciteSetBstMidEndSepPunct{\mcitedefaultmidpunct}
{\mcitedefaultendpunct}{\mcitedefaultseppunct}\relax
\EndOfBibitem
\bibitem[Ramaswamy(2017)]{ramaswamy2017active}
S.~Ramaswamy, \emph{J. Stat. Mech.}, 2017, \textbf{2017}, 054002\relax
\mciteBstWouldAddEndPuncttrue
\mciteSetBstMidEndSepPunct{\mcitedefaultmidpunct}
{\mcitedefaultendpunct}{\mcitedefaultseppunct}\relax
\EndOfBibitem
\bibitem[Gompper \emph{et~al.}(2020)Gompper, Winkler, Speck, Solon, Nardini,
  Peruani, L{\"o}wen, Golestanian, Kaupp,
  Alvarez,\emph{et~al.}]{gompper20202020}
G.~Gompper, R.~G. Winkler, T.~Speck, A.~Solon, C.~Nardini, F.~Peruani,
  H.~L{\"o}wen, R.~Golestanian, U.~B. Kaupp, L.~Alvarez \emph{et~al.}, \emph{J.
  Phys.: Condens. Matter}, 2020, \textbf{32}, 193001\relax
\mciteBstWouldAddEndPuncttrue
\mciteSetBstMidEndSepPunct{\mcitedefaultmidpunct}
{\mcitedefaultendpunct}{\mcitedefaultseppunct}\relax
\EndOfBibitem
\bibitem[Jiang \emph{et~al.}(2010)Jiang, Yoshinaga, and Sano]{jiang2010active}
H.-R. Jiang, N.~Yoshinaga and M.~Sano, \emph{Phys. Rev. Lett.}, 2010,
  \textbf{105}, 268302\relax
\mciteBstWouldAddEndPuncttrue
\mciteSetBstMidEndSepPunct{\mcitedefaultmidpunct}
{\mcitedefaultendpunct}{\mcitedefaultseppunct}\relax
\EndOfBibitem
\bibitem[Romanczuk \emph{et~al.}(2012)Romanczuk, B{\"a}r, Ebeling, Lindner, and
  Schimansky-Geier]{romanczuk2012active}
P.~Romanczuk, M.~B{\"a}r, W.~Ebeling, B.~Lindner and L.~Schimansky-Geier,
  \emph{Eur. Phys. J. Special Topics}, 2012, \textbf{202}, 1\relax
\mciteBstWouldAddEndPuncttrue
\mciteSetBstMidEndSepPunct{\mcitedefaultmidpunct}
{\mcitedefaultendpunct}{\mcitedefaultseppunct}\relax
\EndOfBibitem
\bibitem[Solon \emph{et~al.}(2015)Solon, Cates, and Tailleur]{solon2015active}
A.~P. Solon, M.~E. Cates and J.~Tailleur, \emph{Eur. Phys. J. Special Topics},
  2015, \textbf{224}, 1231\relax
\mciteBstWouldAddEndPuncttrue
\mciteSetBstMidEndSepPunct{\mcitedefaultmidpunct}
{\mcitedefaultendpunct}{\mcitedefaultseppunct}\relax
\EndOfBibitem
\bibitem[Howse \emph{et~al.}(2007)Howse, Jones, Ryan, Gough, Vafabakhsh, and
  Golestanian]{howse2007self}
J.~R. Howse, R.~A. Jones, A.~J. Ryan, T.~Gough, R.~Vafabakhsh and
  R.~Golestanian, \emph{Phys. Rev. Lett.}, 2007, \textbf{99}, 048102\relax
\mciteBstWouldAddEndPuncttrue
\mciteSetBstMidEndSepPunct{\mcitedefaultmidpunct}
{\mcitedefaultendpunct}{\mcitedefaultseppunct}\relax
\EndOfBibitem
\bibitem[Sevilla(2016)]{sevilla2016diffusion}
F.~J. Sevilla, \emph{Phys. Rev. E}, 2016, \textbf{94}, 062120\relax
\mciteBstWouldAddEndPuncttrue
\mciteSetBstMidEndSepPunct{\mcitedefaultmidpunct}
{\mcitedefaultendpunct}{\mcitedefaultseppunct}\relax
\EndOfBibitem
\bibitem[Van~Teeffelen and L{\"o}wen(2008)]{van2008dynamics}
S.~Van~Teeffelen and H.~L{\"o}wen, \emph{Phys. Rev. E}, 2008, \textbf{78},
  020101\relax
\mciteBstWouldAddEndPuncttrue
\mciteSetBstMidEndSepPunct{\mcitedefaultmidpunct}
{\mcitedefaultendpunct}{\mcitedefaultseppunct}\relax
\EndOfBibitem
\bibitem[K{\"u}mmel \emph{et~al.}(2013)K{\"u}mmel, Ten~Hagen, Wittkowski,
  Buttinoni, Eichhorn, Volpe, L{\"o}wen, and Bechinger]{kummel2013circular}
F.~K{\"u}mmel, B.~Ten~Hagen, R.~Wittkowski, I.~Buttinoni, R.~Eichhorn,
  G.~Volpe, H.~L{\"o}wen and C.~Bechinger, \emph{Phys. Rev. Lett.}, 2013,
  \textbf{110}, 198302\relax
\mciteBstWouldAddEndPuncttrue
\mciteSetBstMidEndSepPunct{\mcitedefaultmidpunct}
{\mcitedefaultendpunct}{\mcitedefaultseppunct}\relax
\EndOfBibitem
\bibitem[Kr{\"u}ger \emph{et~al.}(2016)Kr{\"u}ger, Kl{\"o}s, Bahr, and
  Maass]{kruger2016curling}
C.~Kr{\"u}ger, G.~Kl{\"o}s, C.~Bahr and C.~C. Maass, \emph{Phys. Rev. Lett.},
  2016, \textbf{117}, 048003\relax
\mciteBstWouldAddEndPuncttrue
\mciteSetBstMidEndSepPunct{\mcitedefaultmidpunct}
{\mcitedefaultendpunct}{\mcitedefaultseppunct}\relax
\EndOfBibitem
\bibitem[Liebchen and Levis(2022)]{liebchen2022chiral}
B.~Liebchen and D.~Levis, \emph{EPL}, 2022, \textbf{139}, 67001\relax
\mciteBstWouldAddEndPuncttrue
\mciteSetBstMidEndSepPunct{\mcitedefaultmidpunct}
{\mcitedefaultendpunct}{\mcitedefaultseppunct}\relax
\EndOfBibitem
\bibitem[Wykes \emph{et~al.}(2016)Wykes, Palacci, Adachi, Ristroph, Zhong,
  Ward, Zhang, and Shelley]{wykes2016dynamic}
M.~S.~D. Wykes, J.~Palacci, T.~Adachi, L.~Ristroph, X.~Zhong, M.~D. Ward,
  J.~Zhang and M.~J. Shelley, \emph{Soft Matter}, 2016, \textbf{12}, 4584\relax
\mciteBstWouldAddEndPuncttrue
\mciteSetBstMidEndSepPunct{\mcitedefaultmidpunct}
{\mcitedefaultendpunct}{\mcitedefaultseppunct}\relax
\EndOfBibitem
\bibitem[Nosrati \emph{et~al.}(2015)Nosrati, Driouchi, Yip, and
  Sinton]{nosrati2015two}
R.~Nosrati, A.~Driouchi, C.~M. Yip and D.~Sinton, \emph{Nat. Commun.}, 2015,
  \textbf{6}, 1\relax
\mciteBstWouldAddEndPuncttrue
\mciteSetBstMidEndSepPunct{\mcitedefaultmidpunct}
{\mcitedefaultendpunct}{\mcitedefaultseppunct}\relax
\EndOfBibitem
\bibitem[DiLuzio \emph{et~al.}(2005)DiLuzio, Turner, Mayer, Garstecki, Weibel,
  Berg, and Whitesides]{diluzio2005escherichia}
W.~R. DiLuzio, L.~Turner, M.~Mayer, P.~Garstecki, D.~B. Weibel, H.~C. Berg and
  G.~M. Whitesides, \emph{Nature}, 2005, \textbf{435}, 1271\relax
\mciteBstWouldAddEndPuncttrue
\mciteSetBstMidEndSepPunct{\mcitedefaultmidpunct}
{\mcitedefaultendpunct}{\mcitedefaultseppunct}\relax
\EndOfBibitem
\bibitem[Lauga \emph{et~al.}(2006)Lauga, DiLuzio, Whitesides, and
  Stone]{lauga2006swimming}
E.~Lauga, W.~R. DiLuzio, G.~M. Whitesides and H.~A. Stone, \emph{Biophys. J},
  2006, \textbf{90}, 400\relax
\mciteBstWouldAddEndPuncttrue
\mciteSetBstMidEndSepPunct{\mcitedefaultmidpunct}
{\mcitedefaultendpunct}{\mcitedefaultseppunct}\relax
\EndOfBibitem
\bibitem[Narinder \emph{et~al.}(2018)Narinder, Bechinger, and
  Gomez-Solano]{narinder2018memory}
N.~Narinder, C.~Bechinger and J.~R. Gomez-Solano, \emph{Phys. Rev. Lett.},
  2018, \textbf{121}, 078003\relax
\mciteBstWouldAddEndPuncttrue
\mciteSetBstMidEndSepPunct{\mcitedefaultmidpunct}
{\mcitedefaultendpunct}{\mcitedefaultseppunct}\relax
\EndOfBibitem
\bibitem[Crenshaw(1996)]{crenshaw1996new}
H.~C. Crenshaw, \emph{American Zoologist}, 1996, \textbf{36}, 608\relax
\mciteBstWouldAddEndPuncttrue
\mciteSetBstMidEndSepPunct{\mcitedefaultmidpunct}
{\mcitedefaultendpunct}{\mcitedefaultseppunct}\relax
\EndOfBibitem
\bibitem[Ordemann \emph{et~al.}(2003)Ordemann, Balazsi, and
  Moss]{ordemann2003pattern}
A.~Ordemann, G.~Balazsi and F.~Moss, \emph{Physica A}, 2003, \textbf{325},
  260\relax
\mciteBstWouldAddEndPuncttrue
\mciteSetBstMidEndSepPunct{\mcitedefaultmidpunct}
{\mcitedefaultendpunct}{\mcitedefaultseppunct}\relax
\EndOfBibitem
\bibitem[Takagi \emph{et~al.}(2013)Takagi, Braunschweig, Zhang, and
  Shelley]{takagi2013dispersion}
D.~Takagi, A.~B. Braunschweig, J.~Zhang and M.~J. Shelley, \emph{Phys. Rev.
  Lett.}, 2013, \textbf{110}, 038301\relax
\mciteBstWouldAddEndPuncttrue
\mciteSetBstMidEndSepPunct{\mcitedefaultmidpunct}
{\mcitedefaultendpunct}{\mcitedefaultseppunct}\relax
\EndOfBibitem
\bibitem[van Teeffelen \emph{et~al.}(2009)van Teeffelen, Zimmermann, and
  L{\"o}wen]{van2009clockwise}
S.~van Teeffelen, U.~Zimmermann and H.~L{\"o}wen, \emph{Soft Matter}, 2009,
  \textbf{5}, 4510\relax
\mciteBstWouldAddEndPuncttrue
\mciteSetBstMidEndSepPunct{\mcitedefaultmidpunct}
{\mcitedefaultendpunct}{\mcitedefaultseppunct}\relax
\EndOfBibitem
\bibitem[Lancia \emph{et~al.}(2019)Lancia, Yamamoto, Ryabchun, Yamaguchi, Sano,
  and Katsonis]{lancia2019reorientation}
F.~Lancia, T.~Yamamoto, A.~Ryabchun, T.~Yamaguchi, M.~Sano and N.~Katsonis,
  \emph{Nat. Commun.}, 2019, \textbf{10}, 1\relax
\mciteBstWouldAddEndPuncttrue
\mciteSetBstMidEndSepPunct{\mcitedefaultmidpunct}
{\mcitedefaultendpunct}{\mcitedefaultseppunct}\relax
\EndOfBibitem
\bibitem[Ebata and Sano(2015)]{ebata2015swimming}
H.~Ebata and M.~Sano, \emph{Scientific Reports}, 2015, \textbf{5}, 1\relax
\mciteBstWouldAddEndPuncttrue
\mciteSetBstMidEndSepPunct{\mcitedefaultmidpunct}
{\mcitedefaultendpunct}{\mcitedefaultseppunct}\relax
\EndOfBibitem
\bibitem[Tarama and Ohta(2011)]{tarama2011dynamics}
M.~Tarama and T.~Ohta, \emph{Eur. Phys. J. B}, 2011, \textbf{83}, 391\relax
\mciteBstWouldAddEndPuncttrue
\mciteSetBstMidEndSepPunct{\mcitedefaultmidpunct}
{\mcitedefaultendpunct}{\mcitedefaultseppunct}\relax
\EndOfBibitem
\bibitem[Hokmabad \emph{et~al.}(2019)Hokmabad, Baldwin, Kr{\"u}ger, Bahr, and
  Maass]{hokmabad2019topological}
B.~V. Hokmabad, K.~A. Baldwin, C.~Kr{\"u}ger, C.~Bahr and C.~C. Maass,
  \emph{Phys. Rev. Lett.}, 2019, \textbf{123}, 178003\relax
\mciteBstWouldAddEndPuncttrue
\mciteSetBstMidEndSepPunct{\mcitedefaultmidpunct}
{\mcitedefaultendpunct}{\mcitedefaultseppunct}\relax
\EndOfBibitem
\bibitem[Haeggqwist \emph{et~al.}(2008)Haeggqwist, Schimansky-Geier, Sokolov,
  and Moss]{haeggqwist2008hopping}
L.~Haeggqwist, L.~Schimansky-Geier, I.~Sokolov and F.~Moss, \emph{Eur. Phys. J.
  Special Topics}, 2008, \textbf{157}, 33\relax
\mciteBstWouldAddEndPuncttrue
\mciteSetBstMidEndSepPunct{\mcitedefaultmidpunct}
{\mcitedefaultendpunct}{\mcitedefaultseppunct}\relax
\EndOfBibitem
\bibitem[Weber \emph{et~al.}(2011)Weber, Radtke, Schimansky-Geier, and
  H{\"a}nggi]{weber2011active}
C.~Weber, P.~K. Radtke, L.~Schimansky-Geier and P.~H{\"a}nggi, \emph{Phys. Rev.
  E}, 2011, \textbf{84}, 011132\relax
\mciteBstWouldAddEndPuncttrue
\mciteSetBstMidEndSepPunct{\mcitedefaultmidpunct}
{\mcitedefaultendpunct}{\mcitedefaultseppunct}\relax
\EndOfBibitem
\bibitem[Weber \emph{et~al.}(2012)Weber, Sokolov, and
  Schimansky-Geier]{weber2012active}
C.~Weber, I.~M. Sokolov and L.~Schimansky-Geier, \emph{Phys. Rev. E}, 2012,
  \textbf{85}, 052101\relax
\mciteBstWouldAddEndPuncttrue
\mciteSetBstMidEndSepPunct{\mcitedefaultmidpunct}
{\mcitedefaultendpunct}{\mcitedefaultseppunct}\relax
\EndOfBibitem
\bibitem[Olsen(2021)]{olsen2021diffusion}
K.~S. Olsen, \emph{Phys. Rev. E}, 2021, \textbf{103}, 052608\relax
\mciteBstWouldAddEndPuncttrue
\mciteSetBstMidEndSepPunct{\mcitedefaultmidpunct}
{\mcitedefaultendpunct}{\mcitedefaultseppunct}\relax
\EndOfBibitem
\bibitem[Shenoy \emph{et~al.}(2007)Shenoy, Tambe, Prasad, and
  Theriot]{shenoy2007kinematic}
V.~Shenoy, D.~Tambe, A.~Prasad and J.~Theriot, \emph{Proc. Natl. Acad. Sci.},
  2007, \textbf{104}, 8229\relax
\mciteBstWouldAddEndPuncttrue
\mciteSetBstMidEndSepPunct{\mcitedefaultmidpunct}
{\mcitedefaultendpunct}{\mcitedefaultseppunct}\relax
\EndOfBibitem
\bibitem[Malakar \emph{et~al.}(2018)Malakar, Jemseena, Kundu, Kumar,
  Sabhapandit, Majumdar, Redner, and Dhar]{malakar2018steady}
K.~Malakar, V.~Jemseena, A.~Kundu, K.~V. Kumar, S.~Sabhapandit, S.~N. Majumdar,
  S.~Redner and A.~Dhar, \emph{J. Stat. Mech.}, 2018, \textbf{2018},
  043215\relax
\mciteBstWouldAddEndPuncttrue
\mciteSetBstMidEndSepPunct{\mcitedefaultmidpunct}
{\mcitedefaultendpunct}{\mcitedefaultseppunct}\relax
\EndOfBibitem
\bibitem[Basu \emph{et~al.}(2018)Basu, Majumdar, Rosso, and
  Schehr]{basu2018active}
U.~Basu, S.~N. Majumdar, A.~Rosso and G.~Schehr, \emph{Phys. Rev. E}, 2018,
  \textbf{98}, 062121\relax
\mciteBstWouldAddEndPuncttrue
\mciteSetBstMidEndSepPunct{\mcitedefaultmidpunct}
{\mcitedefaultendpunct}{\mcitedefaultseppunct}\relax
\EndOfBibitem
\bibitem[Santra \emph{et~al.}(2021)Santra, Basu, and
  Sabhapandit]{santra2021active}
I.~Santra, U.~Basu and S.~Sabhapandit, \emph{Phys. Rev. E}, 2021, \textbf{104},
  L012601\relax
\mciteBstWouldAddEndPuncttrue
\mciteSetBstMidEndSepPunct{\mcitedefaultmidpunct}
{\mcitedefaultendpunct}{\mcitedefaultseppunct}\relax
\EndOfBibitem
\bibitem[Ebbens \emph{et~al.}(2010)Ebbens, Jones, Ryan, Golestanian, and
  Howse]{ebbens2010self}
S.~Ebbens, R.~A. Jones, A.~J. Ryan, R.~Golestanian and J.~R. Howse, \emph{Phys.
  Rev. E}, 2010, \textbf{82}, 015304\relax
\mciteBstWouldAddEndPuncttrue
\mciteSetBstMidEndSepPunct{\mcitedefaultmidpunct}
{\mcitedefaultendpunct}{\mcitedefaultseppunct}\relax
\EndOfBibitem
\bibitem[Majumdar and Meerson(2020)]{majumdar2020toward}
S.~N. Majumdar and B.~Meerson, \emph{Phys. Rev. E}, 2020, \textbf{102},
  022113\relax
\mciteBstWouldAddEndPuncttrue
\mciteSetBstMidEndSepPunct{\mcitedefaultmidpunct}
{\mcitedefaultendpunct}{\mcitedefaultseppunct}\relax
\EndOfBibitem
\bibitem[Santra \emph{et~al.}(2021)Santra, Basu, and
  Sabhapandit]{santra2021direction}
I.~Santra, U.~Basu and S.~Sabhapandit, \emph{Soft Matter}, 2021, \textbf{17},
  10108\relax
\mciteBstWouldAddEndPuncttrue
\mciteSetBstMidEndSepPunct{\mcitedefaultmidpunct}
{\mcitedefaultendpunct}{\mcitedefaultseppunct}\relax
\EndOfBibitem
\bibitem[Santra \emph{et~al.}(2021)Santra, Das, and Nath]{santra2021brownian}
I.~Santra, S.~Das and S.~K. Nath, \emph{J. Phys. A: Math. Theor.}, 2021,
  \textbf{54}, 334001\relax
\mciteBstWouldAddEndPuncttrue
\mciteSetBstMidEndSepPunct{\mcitedefaultmidpunct}
{\mcitedefaultendpunct}{\mcitedefaultseppunct}\relax
\EndOfBibitem
\end{mcitethebibliography}
\bibliographystyle{crabp_bib}

\end{document}